\documentclass[prb,twocolumn,showpacs,preprintnumbers,amsmath,amssymb]{revtex4}

\usepackage{graphicx}
\usepackage{amsmath}
\usepackage{epstopdf}
\usepackage{amsbsy}
\usepackage{dcolumn}
\usepackage{bm}

\def\k{{\mathbf{k}}}
\def\q{{\mathbf{q}}}
\def\Q{{\mathbf{Q}}}
\def\kQ{{\mathbf{k+Q}}}
\def\mkQ{{\mathbf{-k-Q}}}

\begin{document}

\title{Extinction of quasiparticle interference in underdoped cuprates with coexisting order}
\author{Brian M. Andersen$^1$ and P. J. Hirschfeld$^2$}
\affiliation{$^1$Niels Bohr Institute, University of Copenhagen, Universitetsparken 5, DK-2100 Copenhagen,
Denmark\\
$^2$Department of Physics, University of Florida, Gainesville, Florida 32611-8440, USA}

\date{\today}

\begin{abstract}

Recent scanning tunnelling spectroscopy measurements [Y. Koksaka
{\it et al.}, Nature {\bf 454}, 1072 (2008)] have shown that
dispersing quasiparticle interference peaks  in Fourier
transformed conductance maps disappear as the bias voltage exceeds
a certain threshold corresponding to the coincidence of the
contour of constant quasiparticle  energy with the
antiferromagnetic zone boundary.  Here we argue that this is caused by quasistatic short-range coexisting
order present in the $d$-wave superconducting phase, and that the
most likely origin of this order is disorder-induced incommensurate
antiferromagnetism.  We show explicitly how the peaks are
extinguished in the related situation with coexisting long-range
antiferromagnetic order, and discuss the connection with the
realistic disordered case.   Since  it is the localized quasiparticle interference peaks
rather than the underlying antinodal states themselves which are
destroyed at a critical bias, our proposal resolves a conflict
between scanning tunneling spectroscopy and photoemission regarding the nature of these states.

\end{abstract}


\maketitle

\section{Introduction}

The understanding of how a Mott insulator with localized states
becomes a metal as  one gradually increases the carrier
concentration remains one of the main challenges of condensed
matter physics. This question may be intimately connected with the
so-called nodal-antinodal dichotomy (sharp quasiparticles in the
nodal region versus broad, gapped antinodal "quasiparticles")
observed by angular resolved photoemission spectroscopy (ARPES) on
underdoped cuprate materials.\cite{Damascelli03,FuLee06}
Tunnelling spectroscopy has also been used to probe states in
different regions of momentum space by the help of Fourier transform
scanning tunneling spectroscopy
(FT-STS).\cite{Hoffman1,Howald,Hoffman2,vershinen,mcelroy1,hashimoto} In
particular, it has been argued that the apparent incoherent antinodal
states have their origin in the emergence of charge ordered
regions in the underdoped regime.\cite{mcelroy1}

In the $d$-wave superconducting (dSC) state near optimal doping, quasiparticle
interference (QPI) observed by FT-STS is dominated by peaks at
well-defined wavevectors ${\mathbf{q}}_i$ which agree well with
those predicted by the so-called octet
model.\cite{Hoffman1,mcelroy1,WangLee} The contours of constant
energy (CCE) of the Bogoliubov quasiparticle dispersion are
generally shaped like curved ellipses (``bananas") centered at the
nodal points; seven different nonzero wavevectors ${\mathbf{q}}_i$ connect
the tips of these bananas. The dispersion of the peaks at
${\mathbf{q}}_i$ allows one to extract the shape of the underlying
Fermi surface\cite{Sprunger} as well as the momentum dependence of
the superconducting energy gap.\cite{Hoffman1} A quantitative
understanding of the amplitude and width of the peaks, however, is
not straightforward to obtain, and depends rather sensitively on
the nature of the scattering medium. A point-like scatterer, for
example, in an otherwise homogeneous dSC leads
to a landscape of some spot-like and some arc-like dispersive
features close to ${\mathbf{q}}_i$ in the FT-STS
images\cite{WangLee,Tingfourier,Franzold,capriotti,ZAH04,misra}
whereas experimental data appear mostly spot-like. Furthermore, the weights of the various peaks
calculated in simple theories differ from experiment. There have been
several theoretical attempts to remedy this situation by including
more realistic models for the disorder present in
Bi$_2$Sr$_2$CaCu$_2$O$_{8+\delta}$ (BSCCO). In particular, as
realized recently, it appears important to include the gap
inhomogeneity arising from the dopant
atoms.\cite{mcelroyscience,NAMH05,andersen06,nunner,DellAnna,ChengSu}
Models have also included the extended Coulomb potential arising
in this material from partly screened Bi$\leftrightarrow$Sr
substitutional disorder and the oxygen dopant atoms.\cite{nunner} Nevertheless,
a complete quantitative description of the FT-STS patterns is
still lacking in the superconducting state.

FT-STS has also been used to probe the pseudogap state in the
underdoped regime where non-dispersive (bias-independent)
quasi-periodic conductance modulations has been identified both in
BSCCO\cite{Howald,vershinen,mcelroy1,hashimoto,liu} and
Ca$_{2-x}$Na$_x$CuO$_2$Cl$_2$ (Na-CCOC).\cite{hanaguri1} The
origin of these peaks remains unknown at present but may be caused
by short-range charge order, possibly connected to the existence of nested segments of the Fermi surface near the antinodal regions of the Brillouin zone.\cite{wise} In BSCCO near optimal doping non-dispersive peaks have also been discussed in terms of pinned and disorder-induced charge order.\cite{sachdev,chenyeh,podolsky,AHB03,linda}
In Na-CCOC it has been proposed that phonons play a crucial
role in stabilizing a $d$-wave charge density wave
order,\cite{DHLee} or a surface transition to a commensurate
charge density wave state.\cite{brown} At present, however, it
remains controversial whether true charge ordering is required
for describing the non-dispersive LDOS
modulations.\cite{ghosal,chatterjee,bascones}

Recently, new developments in the FT-STS technique allowed for
further detailed exploration of the electronic properties of
underdoped Na-CCOC and BSCCO. For example, it was argued that
tip-elevation errors can be avoided by studying the
conductance-ratio $Z({\mathbf{r}},E=eV)=g({\mathbf{r}},V)/g({\mathbf{r}},-V)$ where $V$ is the bias
voltage and $g$ the conductance, and the detailed properties of
the LDOS modulations were investigated in this regime as
well.\cite{hanaguri2,kohsaka,hanaguri3} It was found that
irrespective of the doping level, an "extinction line" exists in
momentum space, beyond which most of the dispersing FT-STS peaks
(${\mathbf{q}}_2$,${\mathbf{q}}_3$,${\mathbf{q}}_6$,${\mathbf{q}}_7$)
disappear, to be replaced by a reduced set
(${\mathbf{q}}_1^*$,${\mathbf{q}}_5^*$) of roughly non-dispersive
peaks.\cite{kohsaka} This extinction line is doping independent
and coincides at all doping levels with the antiferromagnetic (AF) zone boundary [lines
joining the points $(0,\pm\pi)$ and $(\pm\pi,0)$]. At energies
below the scale $\Delta_0$ where the CCE first touch the AF zone
boundary, the FT-STS response is similar to the dispersing
Bogoliubov quasiparticles obeying the octet model. At energies above $\Delta_0$, the response
becomes highly spatially inhomogeneous and appears dominated by
pseudogap excitations.

One popular picture of the nodal-antinodal dichotomy invokes
intense scattering with momentum transfer near ($\pi,\pi$) which
broadens  states near the antinodal points. The problem with this picture in
the superconducting state, however, is that the phase space for
scattering is smallest at precisely these points of momentum space
because the $d$-wave gap is largest there. Thus Graser {\it et
al.}\cite{graserprb07} calculated the spin fluctuation spectrum
within an RPA-type formalism and used it to determine the lifetime
of states near the node and antinode of a dSC phase. This same
framework produces a good description of the resonant magnetic
response near ($\pi,\pi$) as measured by neutron
scattering.\cite{eschrigreview}  Nevertheless their results, which
were consistent with earlier work on quasiparticle
lifetimes,\cite{Dahmetal1,Quinlan96,Dahmetal2} imply that
inelastic scattering of the conventional itinerant
spin-fluctuation type cannot severely broaden quasiparticle
states in the superconducting state with momenta near the antinodes.
This is of course consistent with ARPES experiments, which find
broad but well-defined antinodal peaks in the superconducting
state of BSCCO.\cite{Damascelli03,campuzano,shi}

One aspect of the physics of some underdoped cuprate
materials which is left out of the conventional spin-fluctuation
scattering analysis is the possibility of additional order
coexisting with the dSC state at low
temperatures.  Static stripe order has been observed in
several high-$T_c$ materials and appears especially pronounced near 1/8 doping;\cite{Tranquadareview}
in the La$_{2-x}$Ba$_x$CuO$_4$ (LBCO) system, for example, charge order appears around 50K and
persists to lower temperatures, where it coexists with spin order.\cite{fujita} In addition, $\mu$SR has consistently reported so-called ``cluster spin
glass" (CSG) signatures of frozen magnetic order  all over the
underdoped cuprate phase diagram of
La$_{2-x}$Sr$_x$CuO$_4$ (LSCO) and BSCCO,\cite{julien,panagopoulos} generally attributed to disorder
present in signficant amounts due to the way in which these
materials are doped.  In both LSCO and BSCCO, the nodal-antinodal
dichotomy is also observed in ARPES measurements.\cite{xjzhou} These observations suggest that quasiparticle scattering from
short-range coexisting order might play an important role in
explaining the extinction of the QPI peaks in the experiment by
Kohsaka {\it et al.}\cite{kohsaka}

There is no complete consensus on the origin of these
ordering phenomena.  One general notion is that disorder can pin
fluctuating order while still reflecting the intrinsic correlations of the pure system.\cite{kivelson}  Whether the level of
disorder present in these intrinsically disordered systems is too
large to justify such an assumption is not clear.
However, various  concrete models of pinned fluctuating stripes have  been
proposed\cite{alvarez05,atkinson05,robertson,delmaestro,kaul,andersen07}
which resemble experiment in qualitative ways. An alternative
starting point to understand the CSG phase assumes that dopants
nucleate droplets of staggered order, which then interfere
constructively to create quasi-long-range
order.\cite{atkinson05,andersen07}

These studies are related to the important question of whether AF correlations coexisting with preformed Cooper pairs are adequate to
explain the pseudogap phase, or whether other types of ordering
phenomena occur instead, or in
addition.\cite{ZWang,DHLee,seo,vojta08} Models of a
disordered AF phase coexisting with $d$-wave pairs
can reproduce a number of known experimental results: the magnetic
correlations  have been shown to protect the nodal
quasiparticles;\cite{atkinson07,andersenkappa}
reproduce the Fermi arcs and nodal-antinodal
dichotomy;\cite{alvarez08} as well as the temperature dependence
of the superfluid density\cite{atkinson07} and thermal
conductivity.\cite{andersenkappa,andersenDresden}

Although  the coexisting order in these systems is
short-range, we know from neutron measurements in underdoped
LSCO\cite{lake} that correlation lengths can be quite long, of
order 100 lattice spacings.  It may therefore not be a bad
starting point for the study of the effect of these correlations
on QPI to assume a long-range ordered state. Here, we propose a
concrete model for this effect by assuming the presence of a
long-range AF state coexisting with
superconductivity, and then investigating the consequences for the
LDOS modulations. Even though the underdoped materials which
exhibit magnetic ordering are characterized by incommensurate
antiferromagnetism [with incommensurate peaks away from
$(\pi,\pi)$], we assume $(\pi,\pi)$ ordering  for simplicity. We
focus on the origin of the extinction line, and not the 
high-energy LDOS modulations which require more realistic disorder
models, possibly including charge ordering. In order to clearly
elucidate the role of the AF order we consider for simplicity a
single point-like scatterer. The results presented below should be
important for understanding future FT-STS modelling using more
realistic disorder configurations.

\section{Formalism}

The FT-STS signal of a disordered dSC has been
discussed rather extensively by theoretical models.\cite{WangLee,Tingfourier,Franzold,capriotti,ZAH04,misra} In the present
case of coexisting AF order, the translational symmetry is broken,
and the formalism is very similar to the $d$-density wave approach
discussed in e.g. Ref. \onlinecite{bena}. The Hamiltonian reads
\begin{eqnarray}
H&=&\sum_{\k \sigma} \left[\epsilon(\k) -\mu\right]c_{\k\sigma}^\dagger c_{\k\sigma}\\
&+& \sum_{\k} \left[ \sum_\sigma \sigma M c_{\kQ\sigma}^\dagger c_{\k\sigma}  + \Delta(\k) c_{\k\uparrow}^\dagger c_{-\k\downarrow}^\dagger \right] + \mbox{H.c.}\nonumber, \label{eq1}
\end{eqnarray}
where $c_{\k\sigma}^\dagger$ creates an electron with momentum
$\k$ and spin $\sigma$, $M$ is the magnetization and
${\mathbf{Q}}=(\pi,\pi)$ is the AF ordering vector. We work in
units where the lattice constant $a=1$. The superconducting
$d$-wave gap function is $\Delta(\k)=\Delta \left( \cos k_x - \cos
k_y \right)/2$, and the quasiparticle dispersion is
$\epsilon(\k)=\epsilon_1(\k)+\epsilon_2(\k)$, where $\epsilon_1(\k)=
-2t \left( \cos k_x + \cos k_y \right)$ and $\epsilon_2(\k)= -4t'
\cos k_x \cos k_y - 2t'' \left( \cos 2k_x + \cos 2k_y \right)$. It
is convenient to split up the normal state band in this form due
to the different symmetry properties of $\epsilon_1(\k)$ and
$\epsilon_2(\k)$ with respect to momentum shifts of the AF ordering
vector $\Q$; $\epsilon_1(\k+\Q)=-\epsilon_1(\k)$ and
$\epsilon_2(\k+\Q)=\epsilon_2(\k)$. In terms of the following
generalized Nambu spinor $\psi_\k^\dagger=\{
c_{\k\uparrow}^\dagger,c_{\kQ\uparrow}^\dagger,c_{-\k\downarrow},c_{\mkQ\downarrow}\}$,
we can write the Hamiltonian in the form
\begin{equation}
H=\sum_\k \psi_\k^\dagger A(\k) \psi_\k,
\end{equation}
where the sum is restricted to the reduced Brillouin zone (RBZ), $|k_x|+|k_y|\leq \pi$, and $A(\k)$ is given by
\begin{widetext}
\begin{eqnarray}
A(\k)= \left( \begin{array}{cccc}
 \epsilon_1(\k)+\epsilon_2(\k)-\mu & M & \Delta(\k) & 0 \\
M &-\epsilon_1(\k)+\epsilon_2(\k)-\mu &  0 & -\Delta(\k)  \\
 \Delta^*(\k) & 0 &-\epsilon_1(\k)-\epsilon_2(\k)+\mu & M  \\
    0 &-\Delta^*(\k) &M &\epsilon_1(\k)-\epsilon_2(\k)+\mu
    \end{array} \right).
\end{eqnarray}
\end{widetext}
The eigenvalues $\pm E_{1,2}(\k)$ of $A(\k)$ are given by
\begin{equation}
E_{1,2}(\k) = \sqrt{\left([\epsilon_2(\k)-\mu] \pm \sqrt{\epsilon_1^2(\k)+M^2}\right)^2+\Delta^2(\k)},\label{Eig}
\end{equation}
which for $\Delta(\k)=0$ reduces to $E_M^{\pm}(\k)
=[\epsilon_2(\k)-\mu] \pm \sqrt{\epsilon_1^2(\k)+M^2}$, and for
$M=0$ reduces to $E(\k) =\pm
\sqrt{[\epsilon_1(\k)+\epsilon_2(\k)-\mu]^2 + \Delta^2(\k)}$. The
Greens function $G_0(\k,i\omega_n)$ of the pure system is obtained
from the equation
\begin{equation}
G_0(\k,i\omega_n)^{-1}=i \omega_n I - A(\k),
\end{equation}
where $I$ denotes the $4\times 4$ identity matrix.

In the presence of an impurity term
\begin{equation}
H_{imp}=\sum_{\k,\k' \in RBZ} \psi_\k^\dagger V(\k,\k') \psi_{\k'},
\end{equation}
the impurity-contribution to the full Greens function
$G(\k,\k',i\omega_n)$ is given by
\begin{equation}
G(\k,\k',i\omega_n)=
G_0(\k,i\omega_n)T(\k,\k',i\omega_n)G_0(\k',i\omega_n),
\end{equation}
where
\begin{eqnarray}
T(\k,\k',i\omega_n) = V(\k,\k') + \\
\sum_{\k''\in RBZ}V(\k,\k'') G_0(\k'',i\omega_n)T(\k'',\k',i\omega_n) \nonumber.
\label{tmatrix}
\end{eqnarray}
For a point-like impurity $V(\k,\k')$ [and $T(\k,\k',i\omega_n)$] become independent of $\k$ and $\k'$. Specifically, a nonmagnetic $\delta$-function scatterer takes the form
\begin{eqnarray}
V(\k,\k')=V\left( \begin{array}{cccc}
1 &1 &0& 0\\
1 & 1 &0& 0\\
0 &0& -1&-1\\
0 &0 &-1& -1
\end{array} \right ).
\end{eqnarray}

The change in the LDOS from the pure phase $\delta N(\q,\omega)$, is given by\cite{bena}
\begin{eqnarray}
\delta N(\q,\omega)&=&\frac{i}{2\pi}\sum_{\k\in
RBZ}g(\k,\q,\omega), \label{g1}
\end{eqnarray}
where $g(\k,\q,\omega)$ is defined as follows. Let $\k'=\k+\q$.
If $\k'$ is in the RBZ, then
\begin{eqnarray}
g(\k,\q,\omega)= \sum_{i=1}^4 [G_{ii}(\k,\k',
s_i\omega)-G^*_{ii}(\k',\k,s_i\omega)],
\end{eqnarray}
where $s_i=1$ for for the particle-hole sector $i=1,2$ and $s_i=-1$ for the hole-particle sector $i=3,4$. If $\k'$ is
not in the RBZ, let $\k''=\k+\q-\Q$. For this case
\begin{eqnarray}
g(\k,\q,\omega) &=&\sum_{i=1,3}
[G_{i,i+1}(\k,\k'',s_i\omega)-G^*_{i,i+1}(\k'',\k,s_i\omega) \nonumber
\\& +&G_{i+1,i}(\k,\k'',s_i\omega)-G^*_{i+1,i}(\k'',\k,s_i\omega)].
\label{g3}
\end{eqnarray}
Here, $G(\k,\k',\omega)$ is obtained by usual analytical
continuation $i \omega_n \rightarrow \omega+i0^+$ of $G(\k,\k',i
\omega_n)$. Below we introduce a finite lifetime broadening
$\eta=0.02t$ such that $i \omega_n \rightarrow \omega+i\eta$, and
the summation over the RBZ is performed using a $600\times 600$
mesh.

\begin{figure}[]
\includegraphics[clip=true,width=1.0\columnwidth]{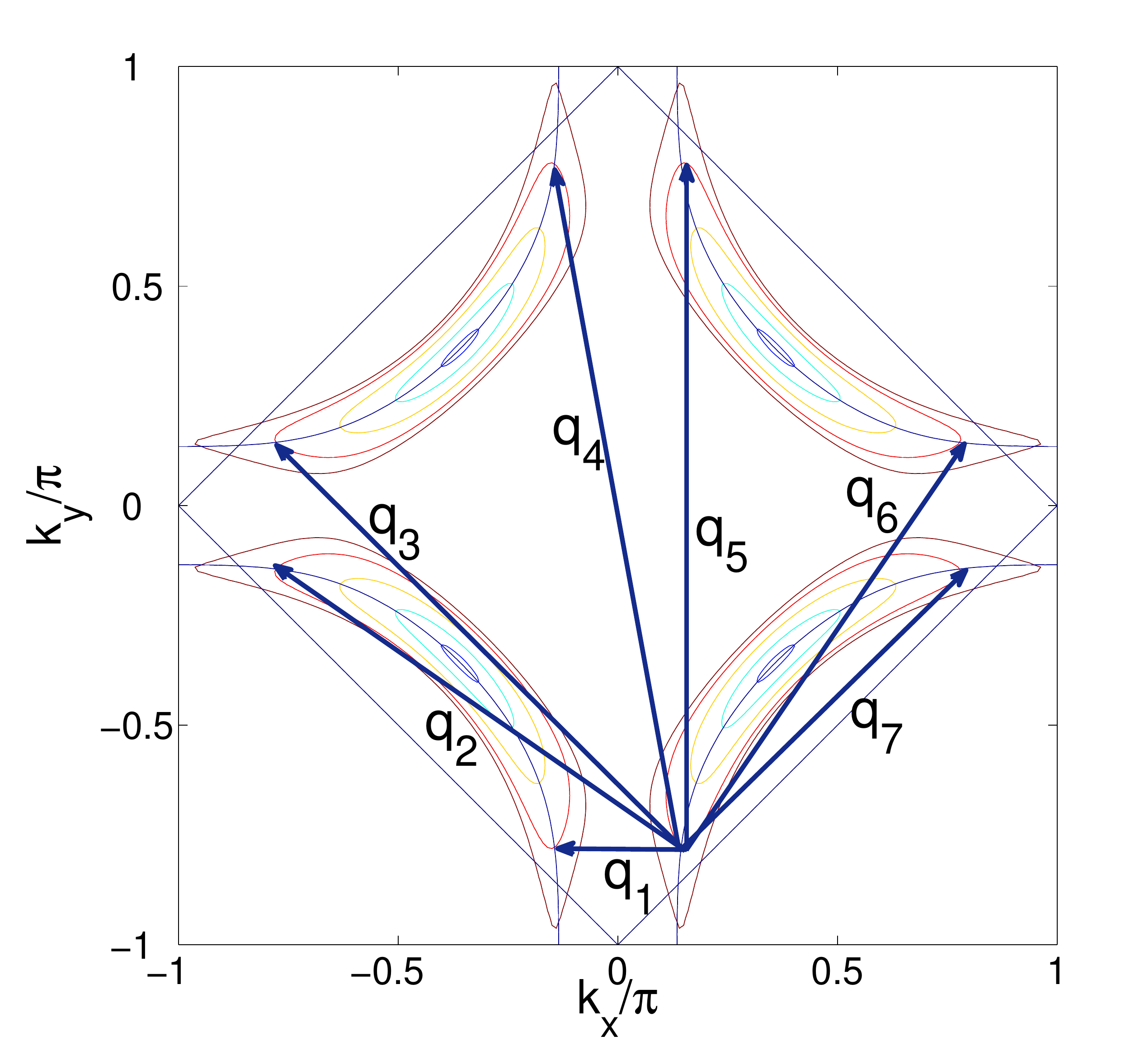}
\caption{(Color online) (a) Contours of constant quasiparticle energy for  the
pure $d$-wave superconductor with $\Delta=0.6t$ at energies
$\omega/t=0.0, 0.075, 0.225, 0.375, 0.5, 0.57$. Also
shown are the underlying normal state Fermi surface, the AF
zone boundary, and the seven distinct nonzero wavevectors ${\bf q}_i$ connecting the banana tips.} \label{fig1}
\end{figure}

We use a band dispersion with $t=1.0, t'=-0.3, t''=0.1$ and
$\mu=-1.25$, which yields the normal state Fermi surface shown in
Fig. \ref{fig1}. In addition, when studying the superconducting
state we set $\Delta=0.6t$. In the following we focus the discussion on energies below this gap energy, which is taken unrealistically
large for numerical purposes in order to highlight features at low energies. The impurity potential $V=0.1t$ is
chosen to be weak in the sense that it does not produce any
low-energy resonant states. For simplicity, we ignore any spatial
structure of the local Wannier orbitals, rendering all results
periodic in momentum space with respect to reciprocal lattice
vectors.

\begin{figure}[t]
\includegraphics[clip=true,height=1.5\columnwidth,width=1.0\columnwidth]{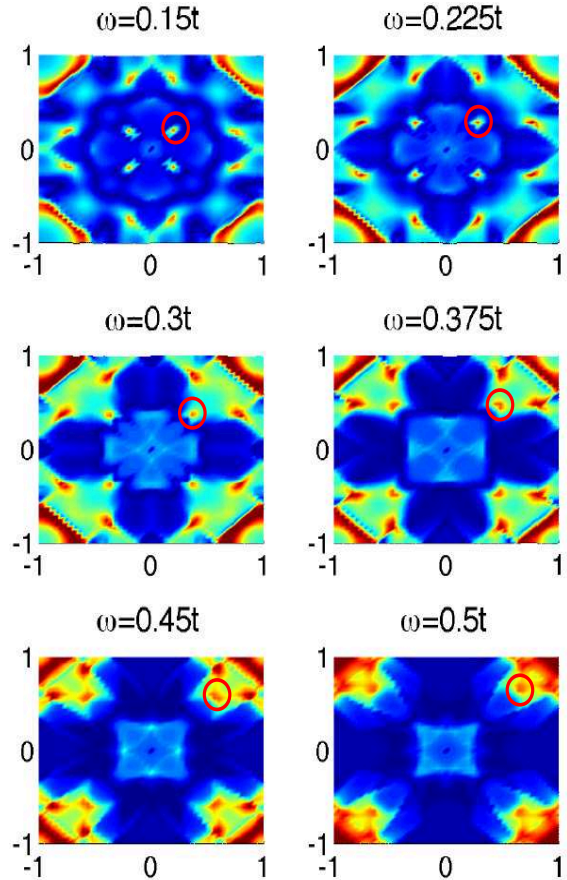}
\caption{(Color online) QPI maps versus $q_x/\pi$ and $q_y/\pi$ in the dSC phase in the presence
of single point-like potential scatterer of strength $V=0.1t$. Energies
$\omega$ are as shown in the Figure titles. Some dispersing
spot-like features corresponding to octet vectors ${\bf q}_i$ are
easy to identify. For example, we have circled the ${\bf q}_7$ peak [see Fig. \ref{fig1}].}\label{fig2}
\end{figure}

\section{Superconducting Phase}

\begin{figure}[t]
\begin{minipage}{.49\columnwidth}
\includegraphics[clip=true,height=0.8\columnwidth,width=0.98\columnwidth]{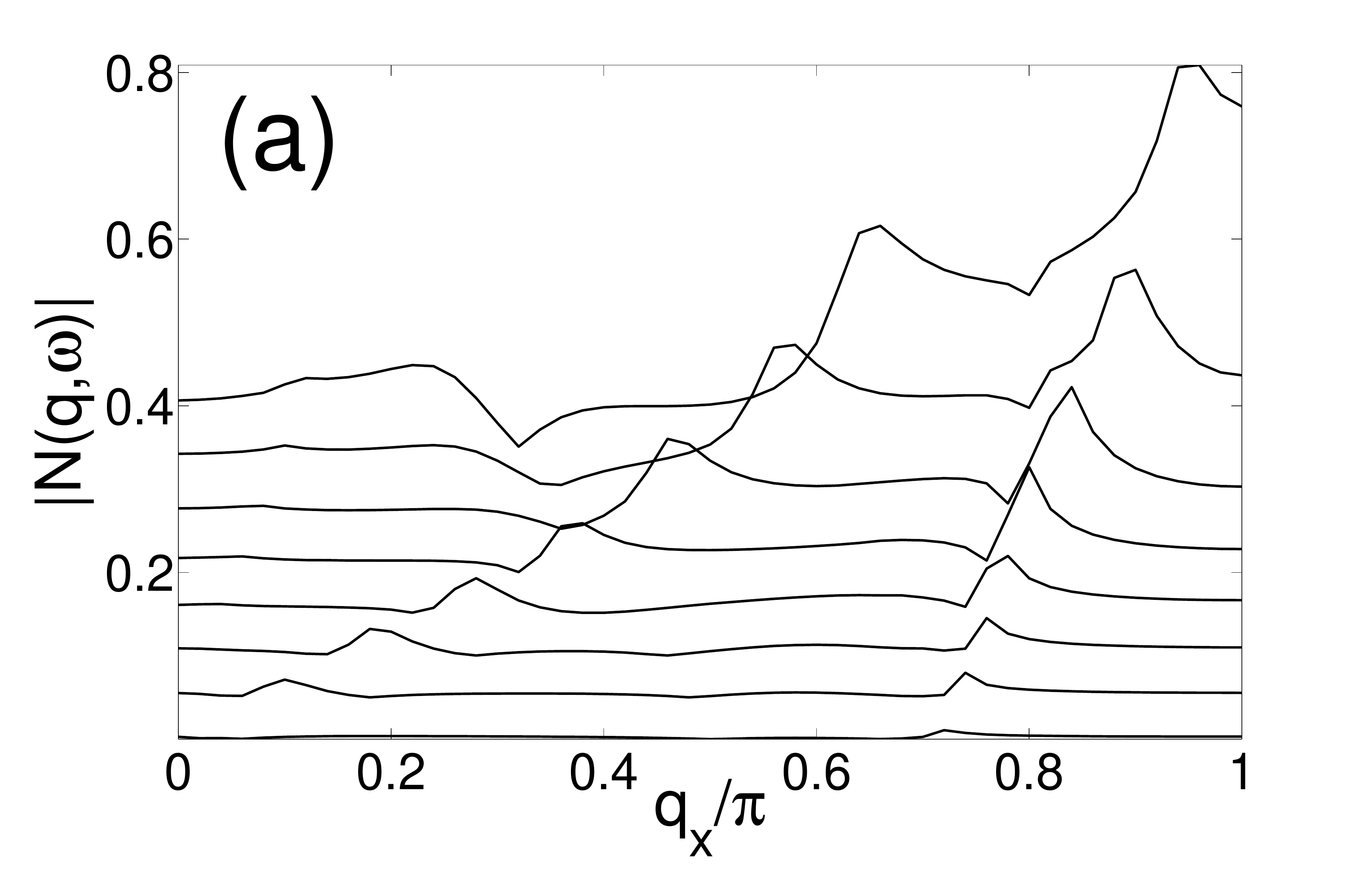}
\end{minipage}
\begin{minipage}{.49\columnwidth}
\includegraphics[clip=true,height=0.8\columnwidth,width=0.98\columnwidth]{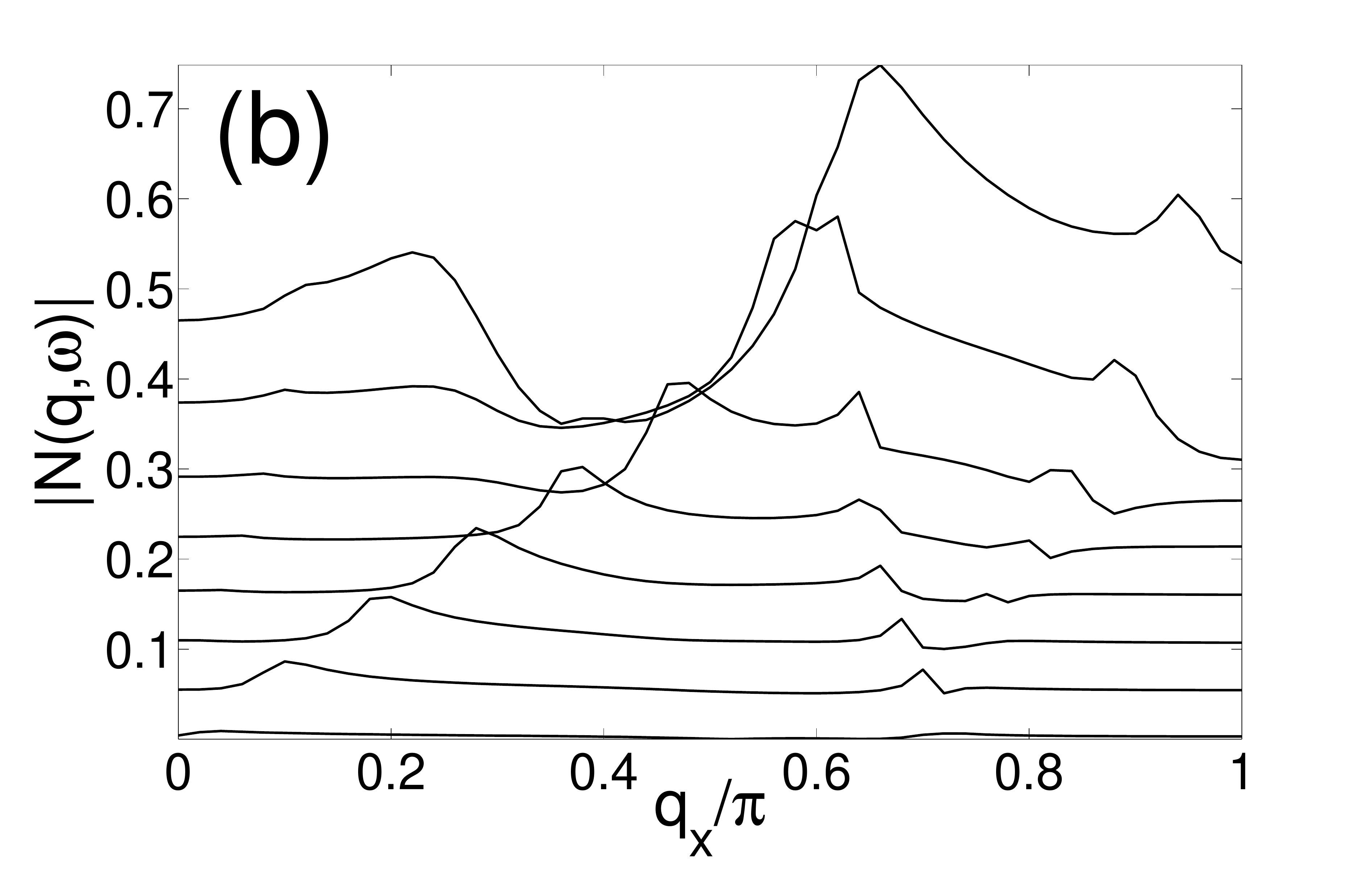}
\end{minipage}
\\
\begin{minipage}{.49\columnwidth}
\includegraphics[clip=true,height=0.8\columnwidth,width=0.98\columnwidth]{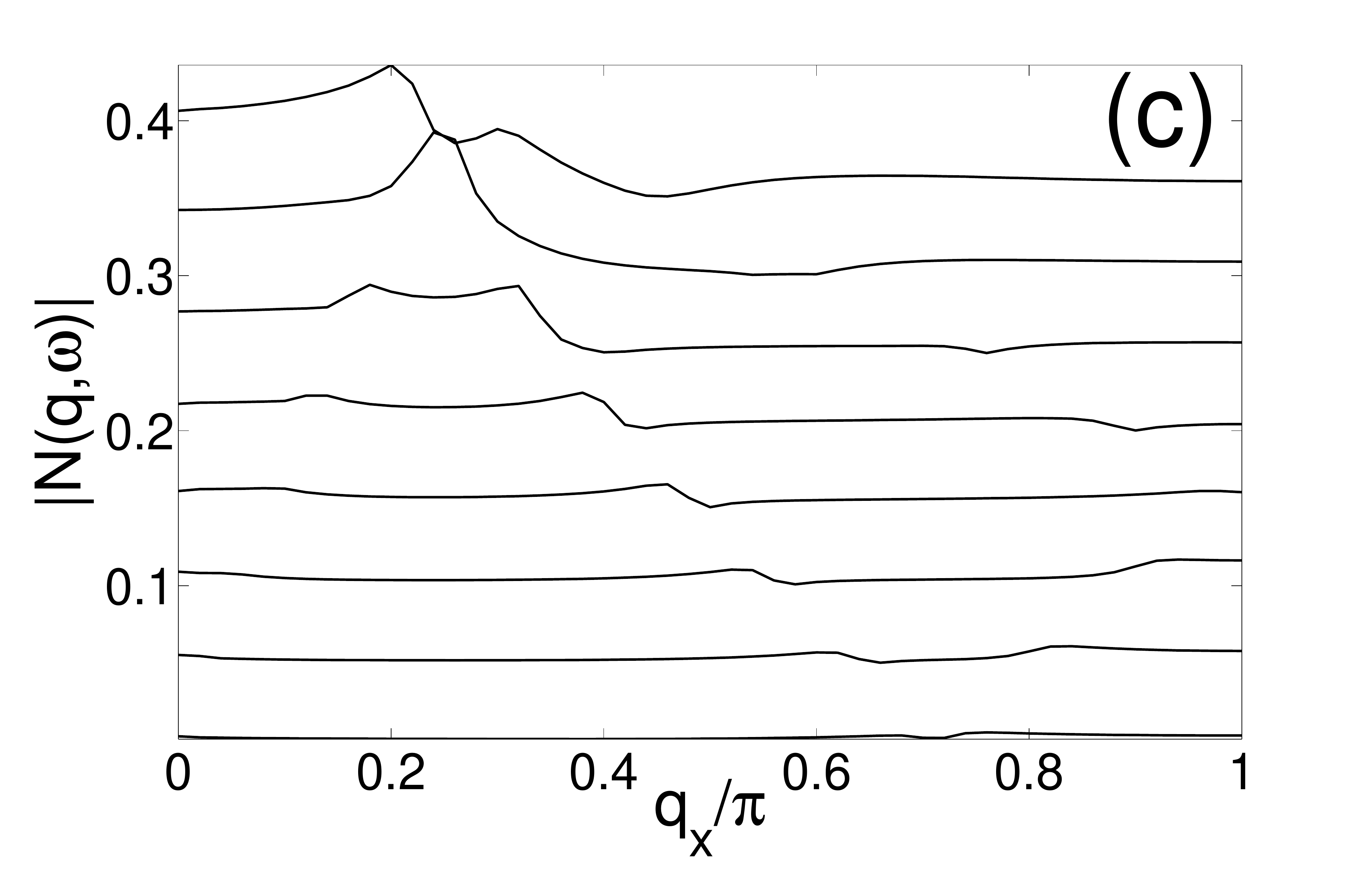}
\end{minipage}
\begin{minipage}{.49\columnwidth}
\includegraphics[clip=true,height=0.8\columnwidth,width=0.98\columnwidth]{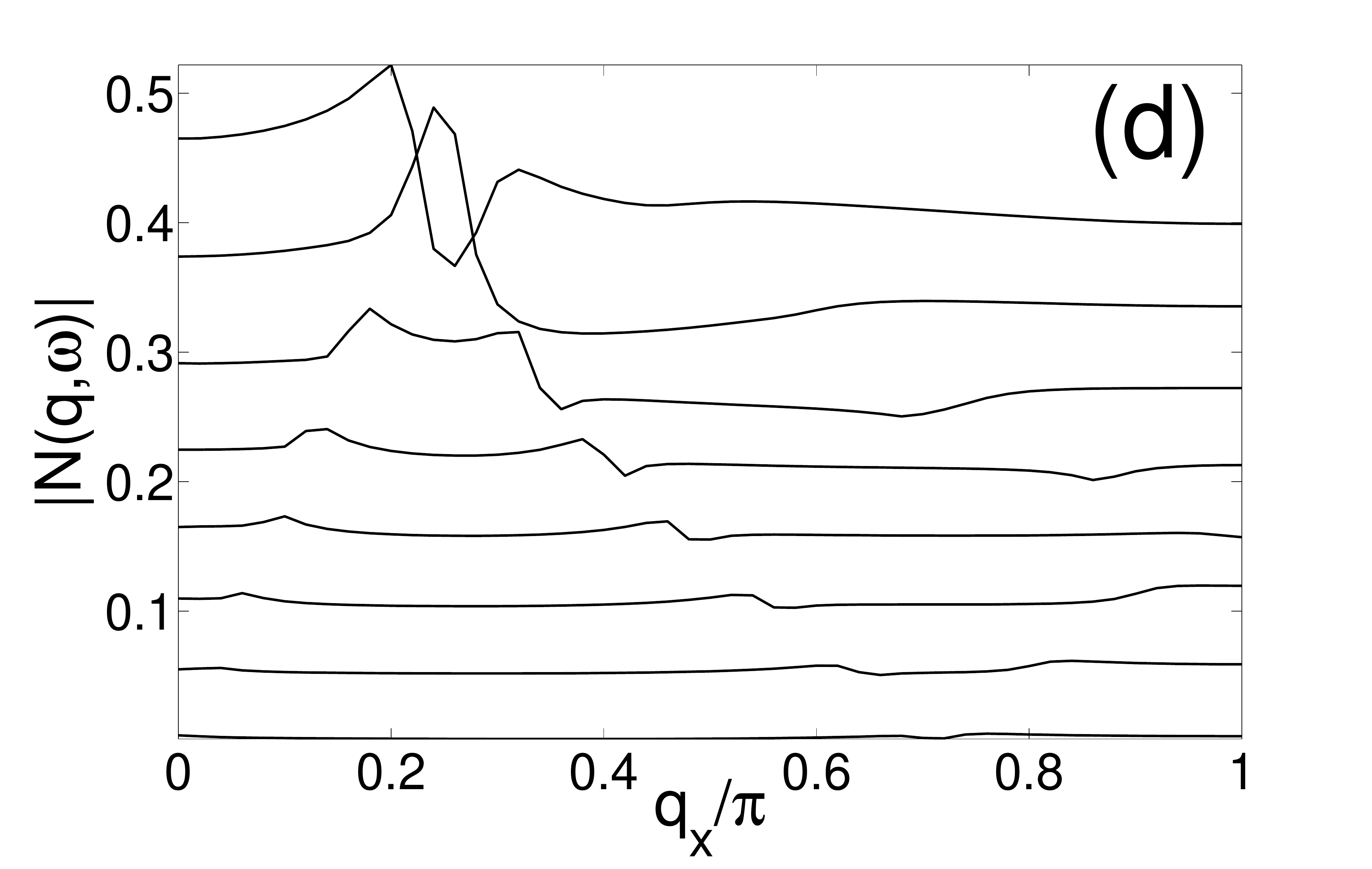}
\end{minipage}
\caption{QPI line cuts along the (110) (a,b) and (100) (c,d)
directions for the pure dSC phase. Panels (a,c) [(b,d)]
correspond to positive [negative] energies at
$|\omega|/t=0.01,0.075,0.15,0.225,0.3,0.375,0.45,0.5$ (bottom to top). For clarity the curves are displaced by 0.05.} \label{fig3}
\end{figure}

The QPI patterns in the pure dSC phase have been thoroughly
discussed in the literature, and we will not
dwell on them here. However, in order to discuss the effect of AF
on the QPI, we show briefly some results for the pure dSC phase in
this section. Figure \ref{fig1} displays some typical CCE exhibiting the usual
banana shaped form, and centered at the nodal points $(\pm 0.36,\pm 0.36)\pi$. The wavevectors ${\mathbf{q}}_i$ connecting
the tips of these bananas, reveal where peaks in the FT-STS maps are expected,
although matrix elements consisting of certain combinations of
coherence factors are important for this simple picture to hold.\cite{WangLee,Franzold} Here
we calculate the Fourier transform density of states $|\delta N|$ with $\delta N$ given by
Eq.(\ref{g1}), and refer to it as a QPI map.   Figure \ref{fig2} shows QPI maps versus $q_x$ and $q_y$ at representative fixed energies
inside the gap. For clarity, we have circled the ${\bf q}_7$ peak which is positioned along the (110) direction and disperses to higher momenta with increasing energy as expected from Fig. \ref{fig1}. Line cuts along the nodal and antinodal directions for the pure dSC phase are shown in Fig. \ref{fig3}
for both positive and negative energies, exhibiting clearly the dispersive QPI peaks
discussed  in the literature.  The dispersion agrees well with the octet
model,\cite{Hoffman1} which assumes that the scattering peaks are
determined entirely by the wavevectors connecting each banana tip with
all the others. It is easy to identify in Fig. \ref{fig3} the octet peaks ${\bf q}_3$ and ${\bf q}_7$
in the (110) cut, and the ${\bf q}_1$ and ${\bf q}_5$ peaks in the (100)
cut. 

\section{Antiferromagnetic Phase}

\begin{figure}[b]
\begin{minipage}{.49\columnwidth}
\includegraphics[clip=true,width=0.98\columnwidth]{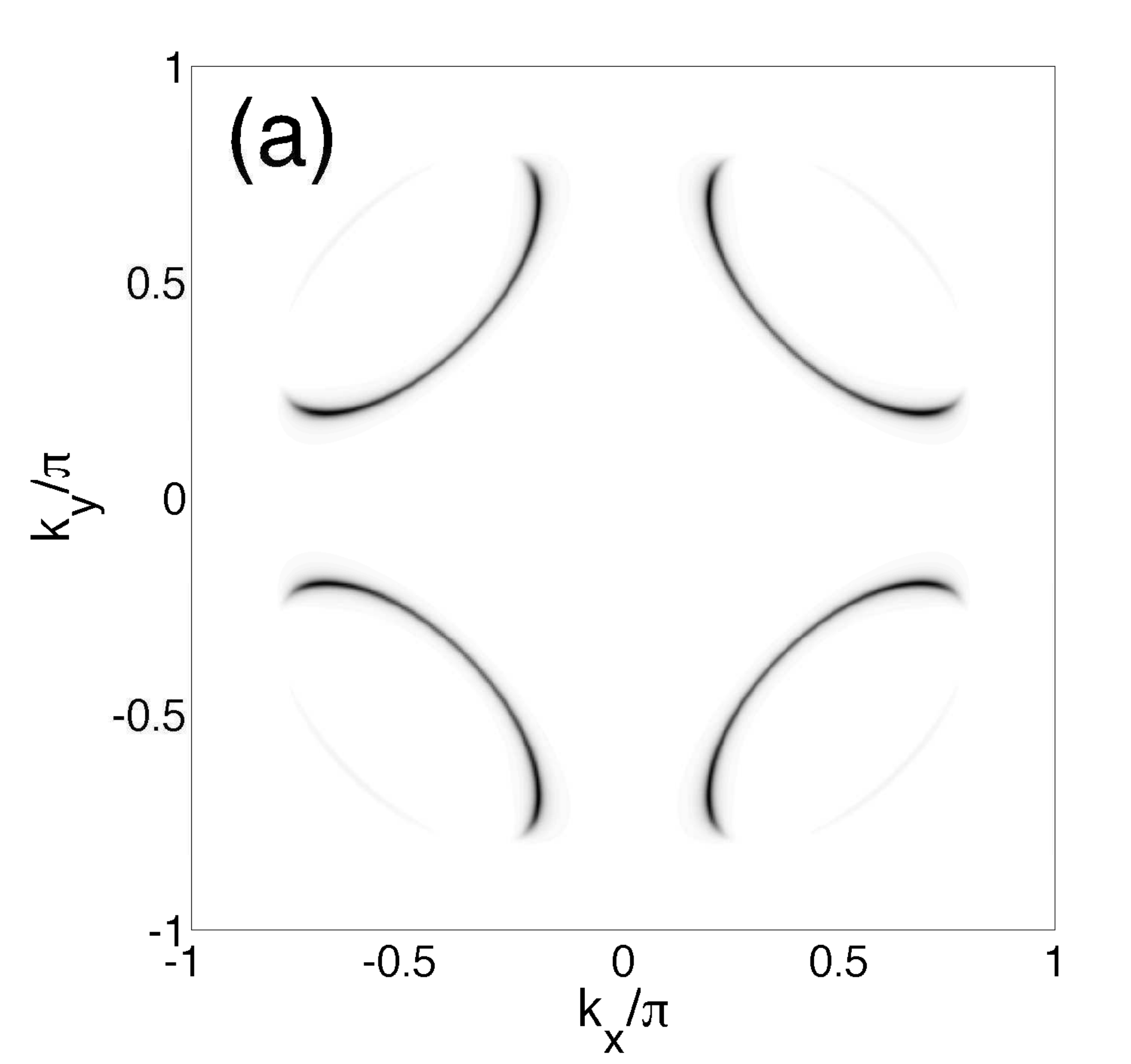}
\end{minipage}
\begin{minipage}{.49\columnwidth}
\includegraphics[clip=true,width=0.98\columnwidth]{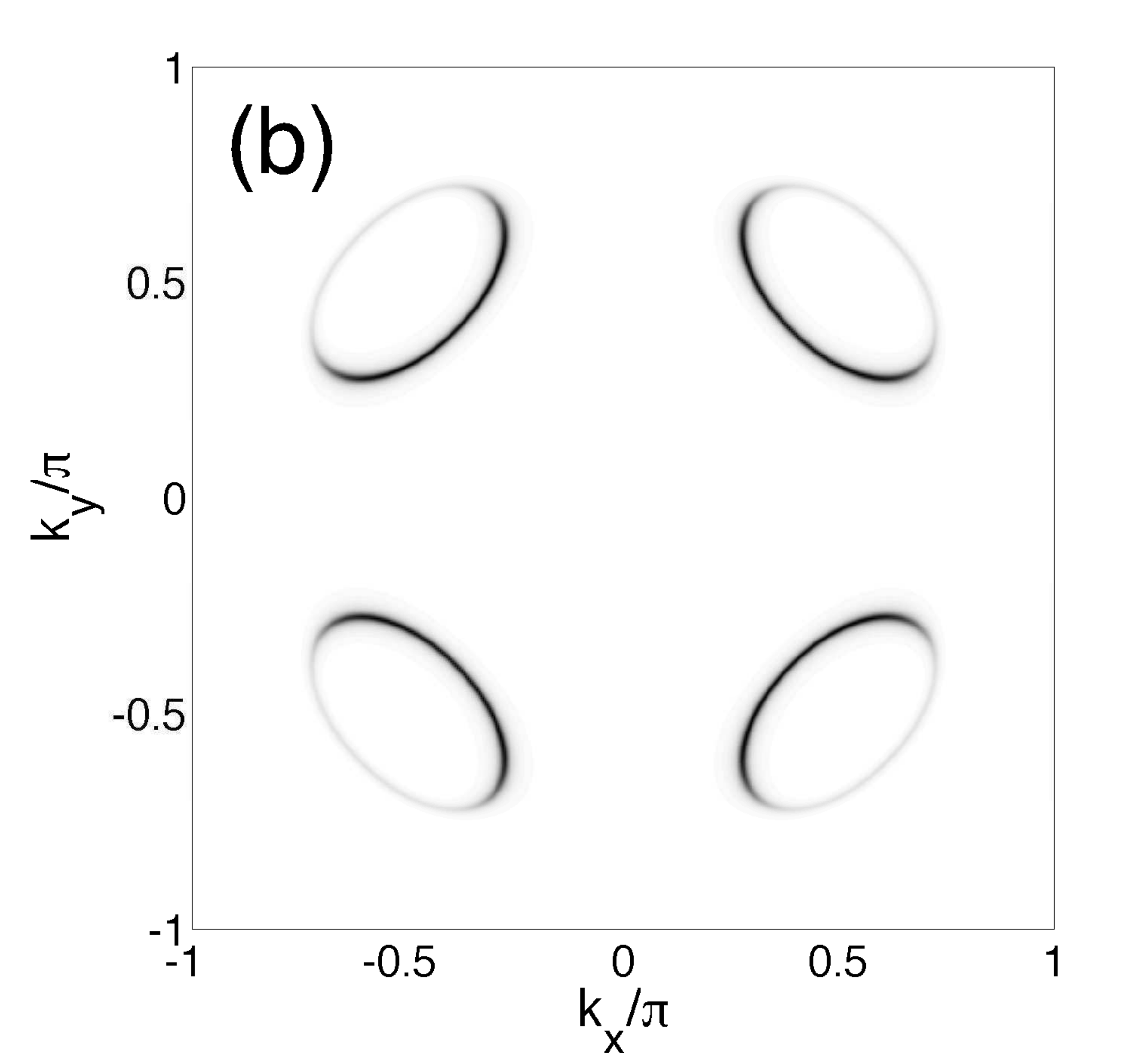}
\end{minipage}
\\
\begin{minipage}{.49\columnwidth}
\includegraphics[clip=true,width=0.98\columnwidth]{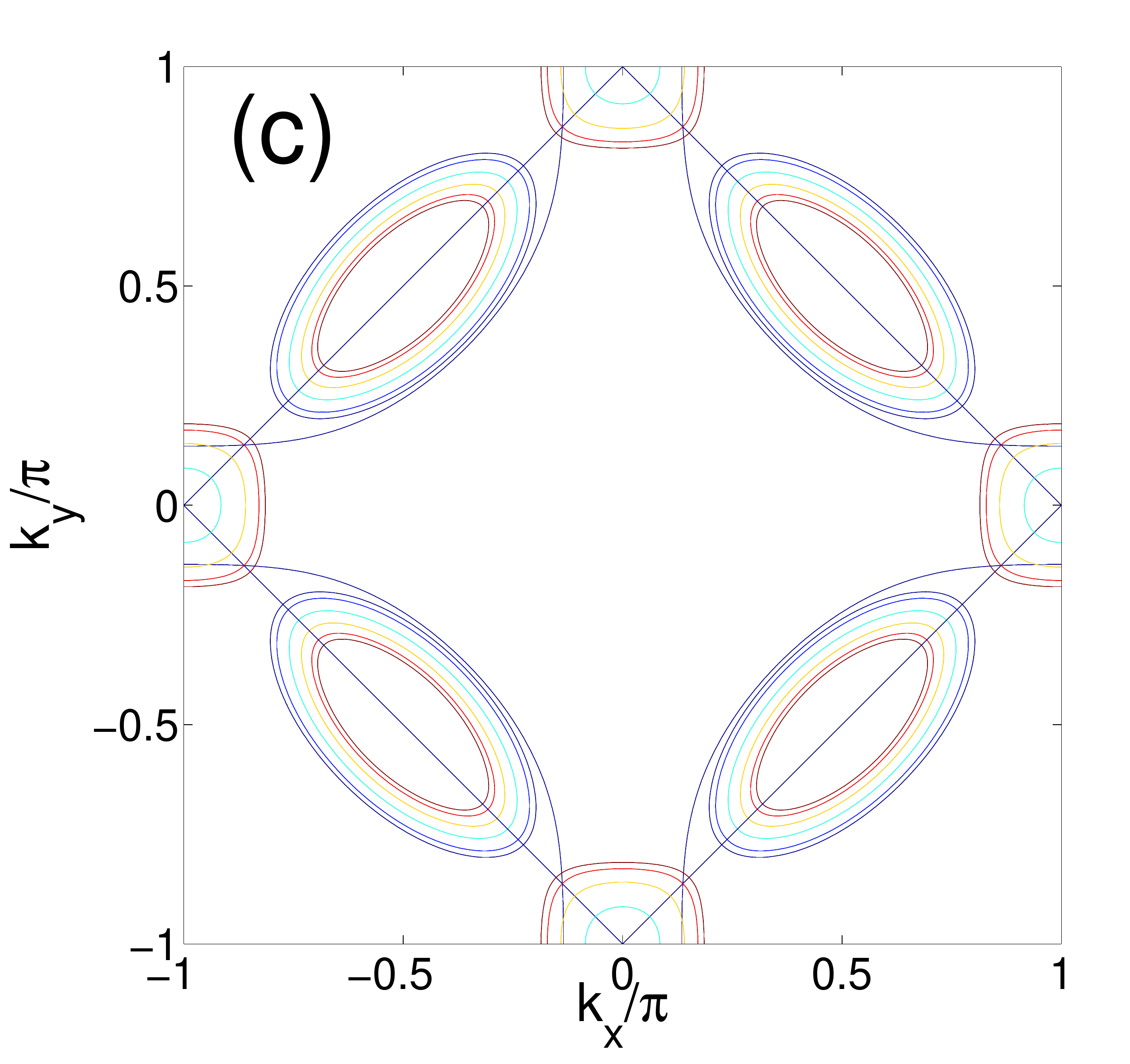}
\end{minipage}
\begin{minipage}{.49\columnwidth}
\includegraphics[clip=true,width=0.98\columnwidth]{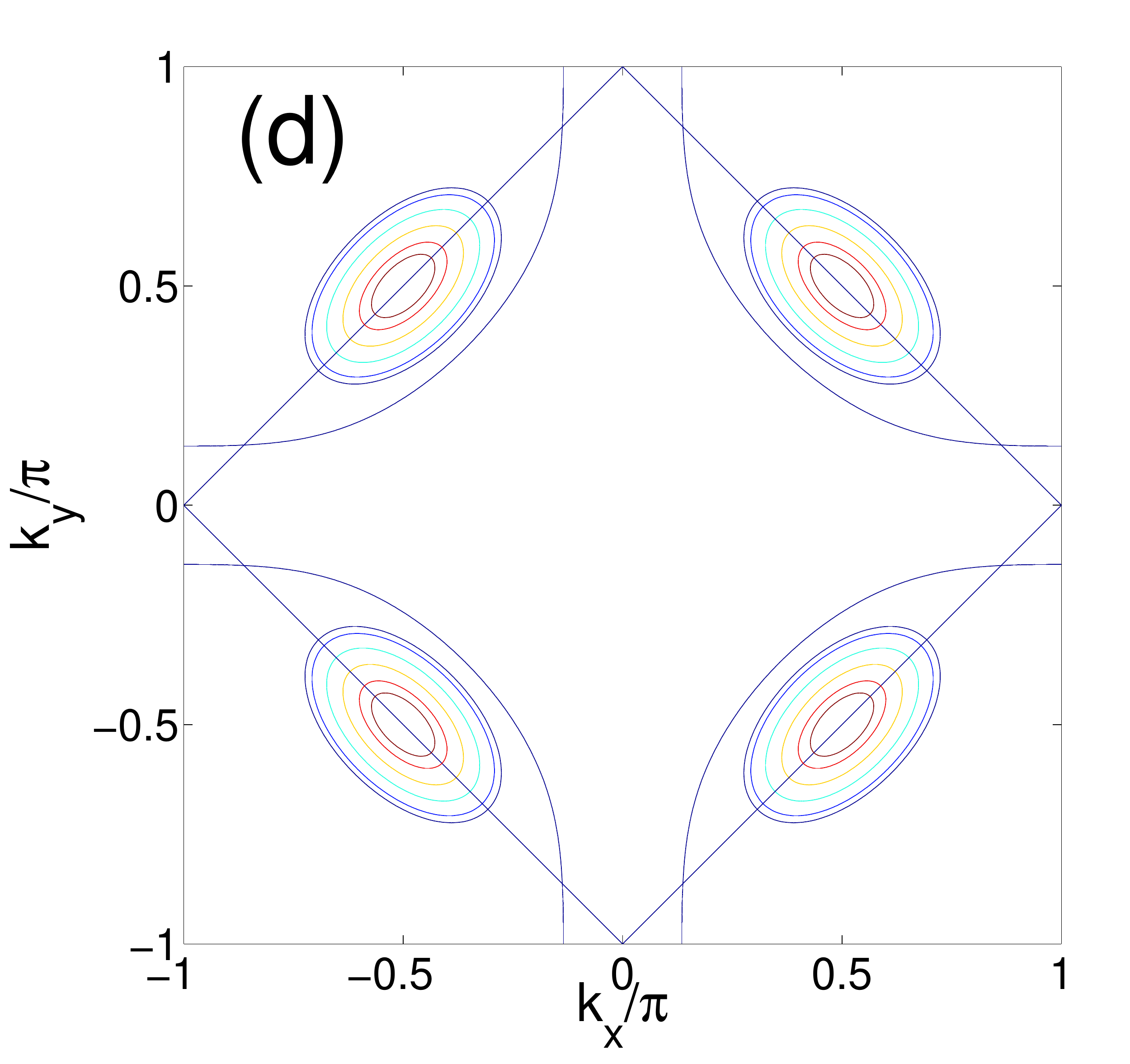}
\end{minipage}
\caption{(Color online) Spectral weight $A(\k,\omega=0)$ in the pure AF
phase with $M=0.5t$ (a), and $M=t$ (b). The lower two panels (c,d)
show CCE at the same energies as in Fig. \ref{fig1} for $M=0.5t$
(c), and $M=t$ (d).} \label{fig4}
\end{figure}

In the pure dSC state, the spectral weight at the Fermi energy $A(\k,\omega=0)$ consists of four nodal points, which may be smeared slightly by disorder. On the other hand, in the pure metallic AF case, the spectral function can be written as
\begin{equation}
A(\k,\omega)=v^2(\k)\delta[\omega-E_M^-(\k)] + u^2(\k)\delta[\omega-E_M^+(\k)],
\end{equation}
with $u^2(\k)=1/2[1+\epsilon_1(\k)/\sqrt{M^2+\epsilon_1^2(\k)}]$ and
$v^2(\k)=1/2[1-\epsilon_1(\k)/\sqrt{M^2+\epsilon_1^2(\k)}]$. In Fig.
\ref{fig4}(a,b) we show the spectral weight $A(\k,\omega=0)$ in a case with
$M=0.5t$ (a) and $M=t$ (b). In both cases only the $E_M^-$ band
crosses the Fermi level. As seen in Fig. \ref{fig4}(a,b), the
outer ring of the Fermi pocket is washed out because of the
$v^2(\k)$ coherence factor. This is caused by the unit cell
doubling in the AF state, and a similar effect happens in e.g. the
$d$-density wave scenario.\cite{chakravarty} The checkerboard
charge-order scenario for the pseudogap phase also reproduces a
Fermi arc due to the difference in the coherence factors between
the inner and outer parts of the arc.\cite{ZWang,DHLee} Disordered
antiferromagnetism will further enhance this apparent spectral
weight suppression on the outside of the pockets.\cite{singleton} 

\begin{figure}[]
\includegraphics[clip=true,height=1.5\columnwidth,width=1.0\columnwidth]{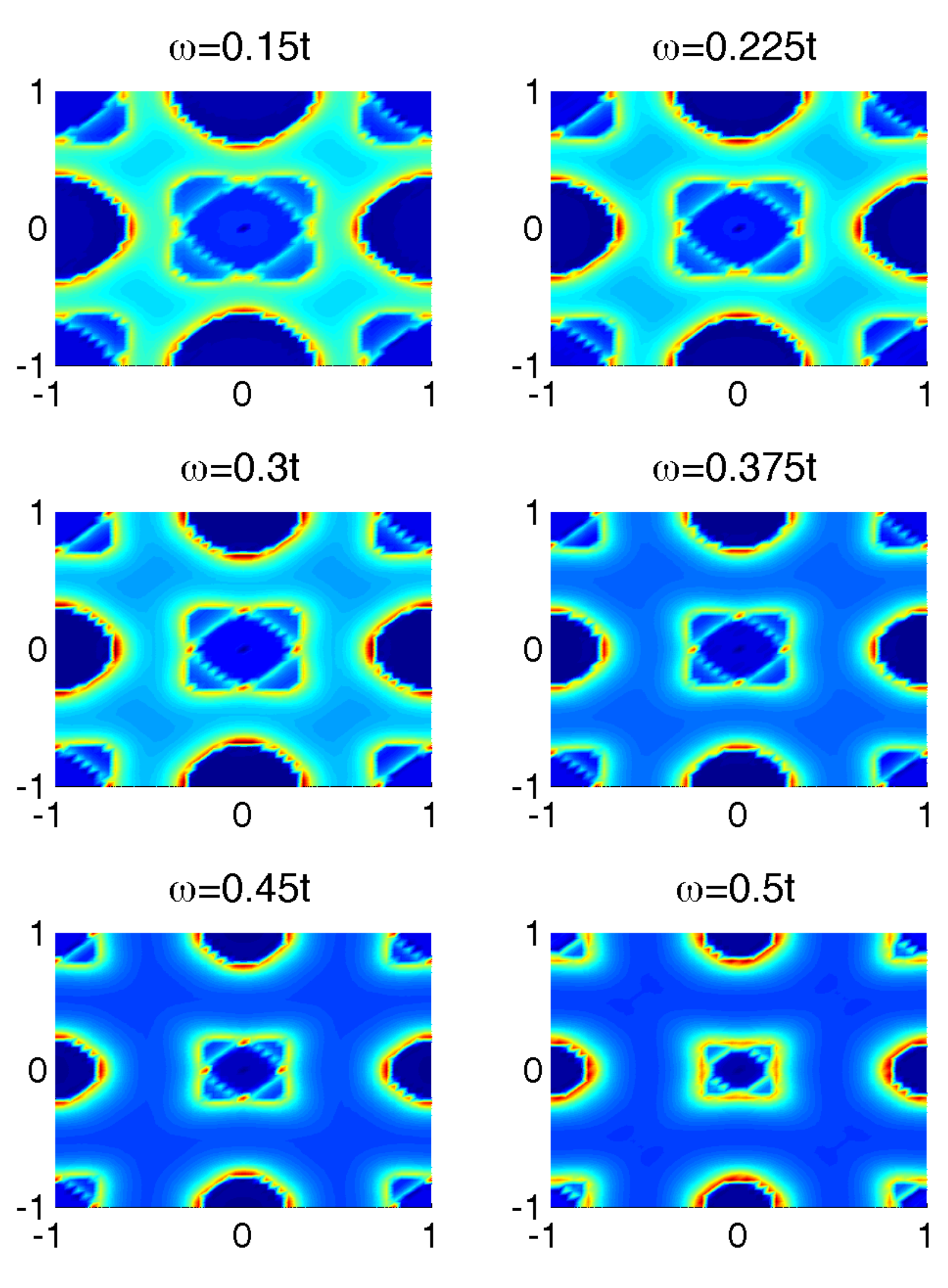}
\caption{(Color online) Same as Fig. \ref{fig2} but for the pure AF phase with $M=t$.} \label{fig5}
\end{figure}

For the case with $M=t$ we show in Fig. \ref{fig5} typical QPI maps in the pure AF phase. The QPI images are dominated by arcs of scattering intensity, rather than spots. 
The reason why distinct, isolated scattering wavevectors do not occur in the pure AF case is caused be the different coherence factors in this case. As discussed in Ref. \onlinecite{Franzold} in the case of intra-nodal scattering, the origin of peaks in the QPI maps in the pure dSC phase is caused by the dSC coherence factors which conspire to significantly enhance the weight near the tips of the CCE resulting is a significant enhancement of the QPI response localized near ${\bf q}_7$. In the pure AF phase, on the other hand, the different coherence factors cause the weight to be evenly distributed along the CCE resulting in arc-like characteristic QPI features. We note that the momentum-resolved density of states, which is qualitatively similar (at least at negative energies) in the two phases, is not the cause of this qualitative difference in the QPI maps. 

\begin{figure}[b]
\begin{minipage}{.49\columnwidth}
\includegraphics[clip=true,height=0.8\columnwidth,width=0.98\columnwidth]{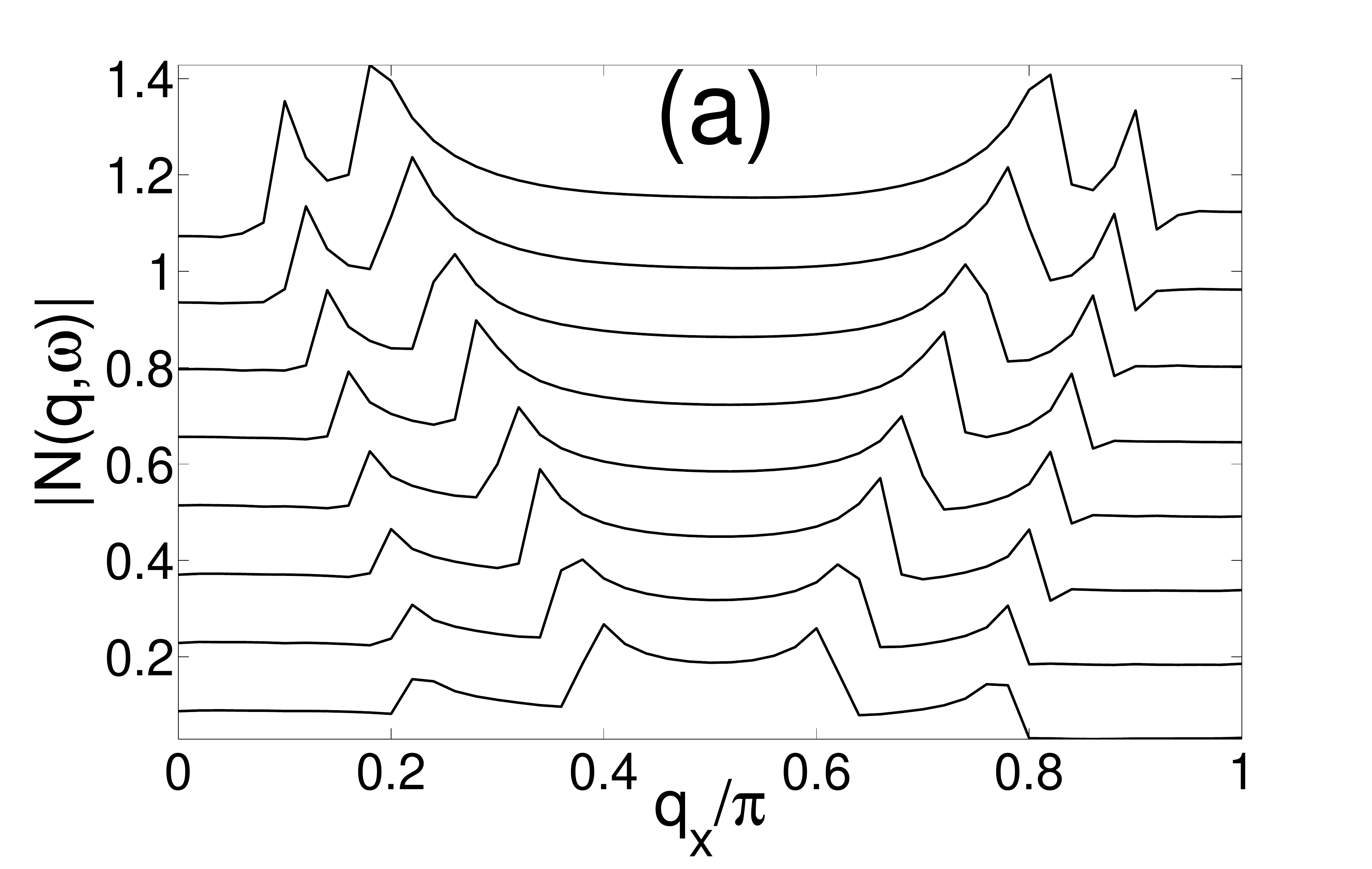}
\end{minipage}
\begin{minipage}{.49\columnwidth}
\includegraphics[clip=true,height=0.8\columnwidth,width=0.98\columnwidth]{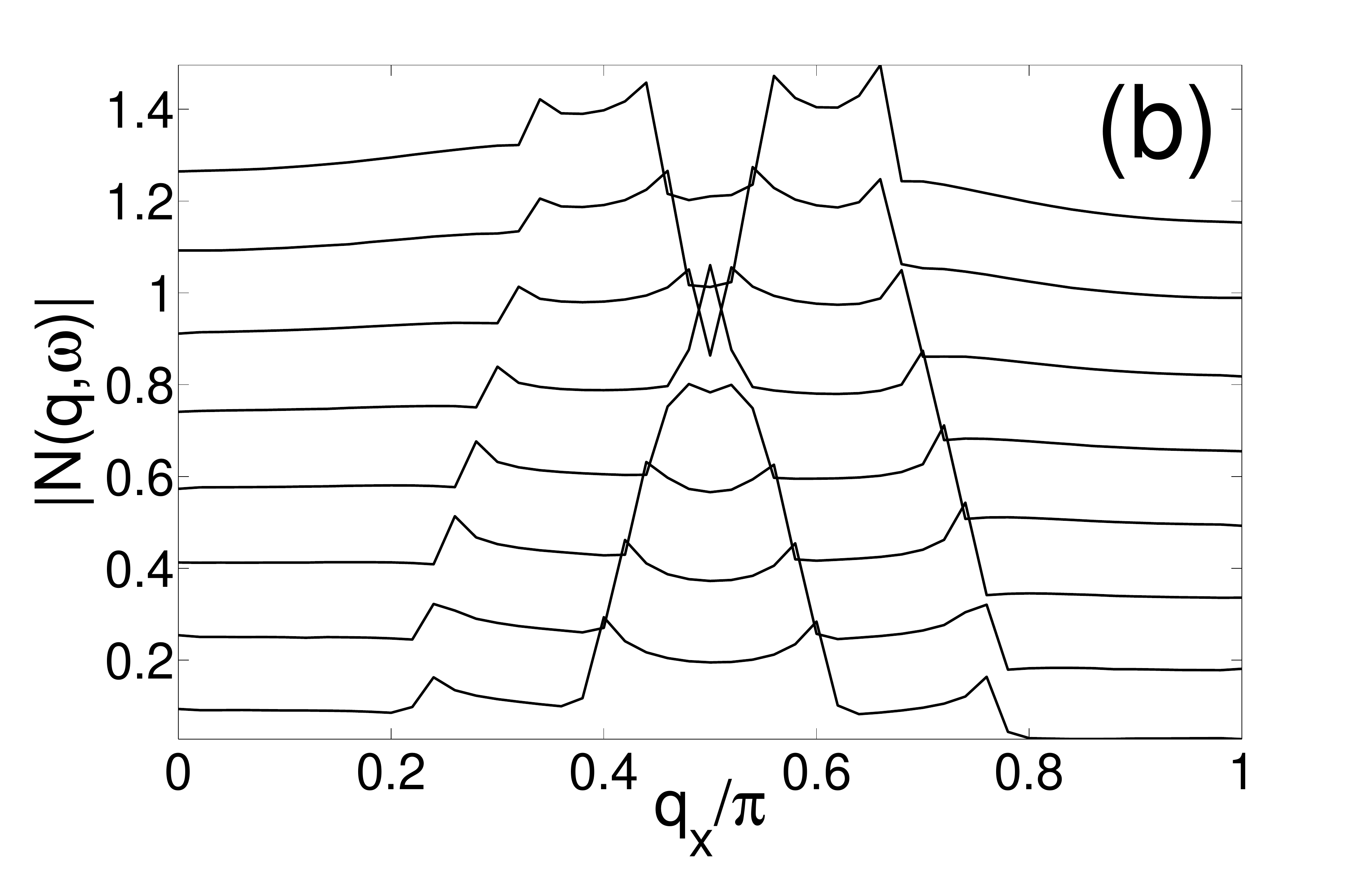}
\end{minipage}
\\
\begin{minipage}{.49\columnwidth}
\includegraphics[clip=true,height=0.8\columnwidth,width=0.98\columnwidth]{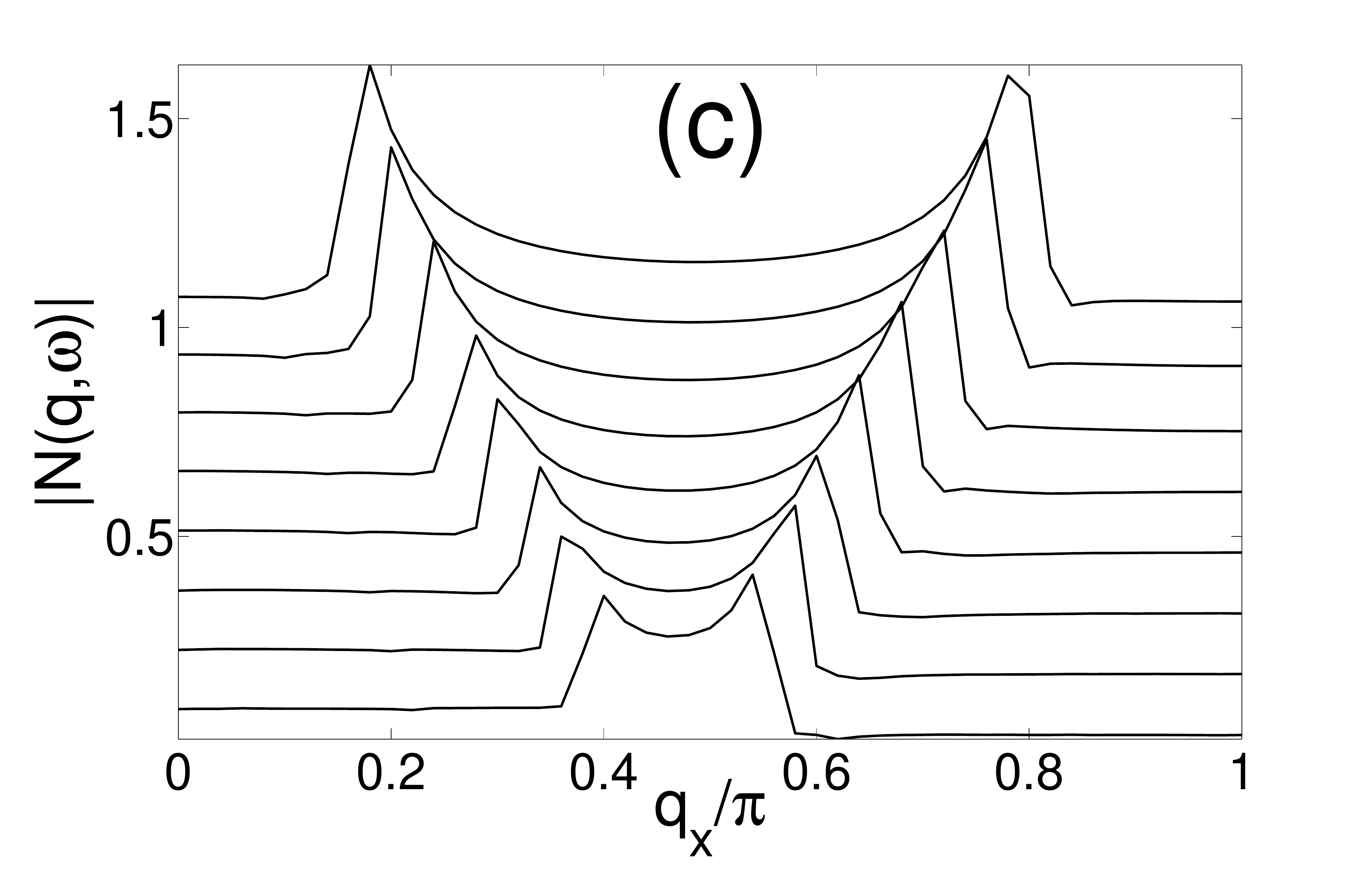}
\end{minipage}
\begin{minipage}{.49\columnwidth}
\includegraphics[clip=true,height=0.8\columnwidth,width=0.98\columnwidth]{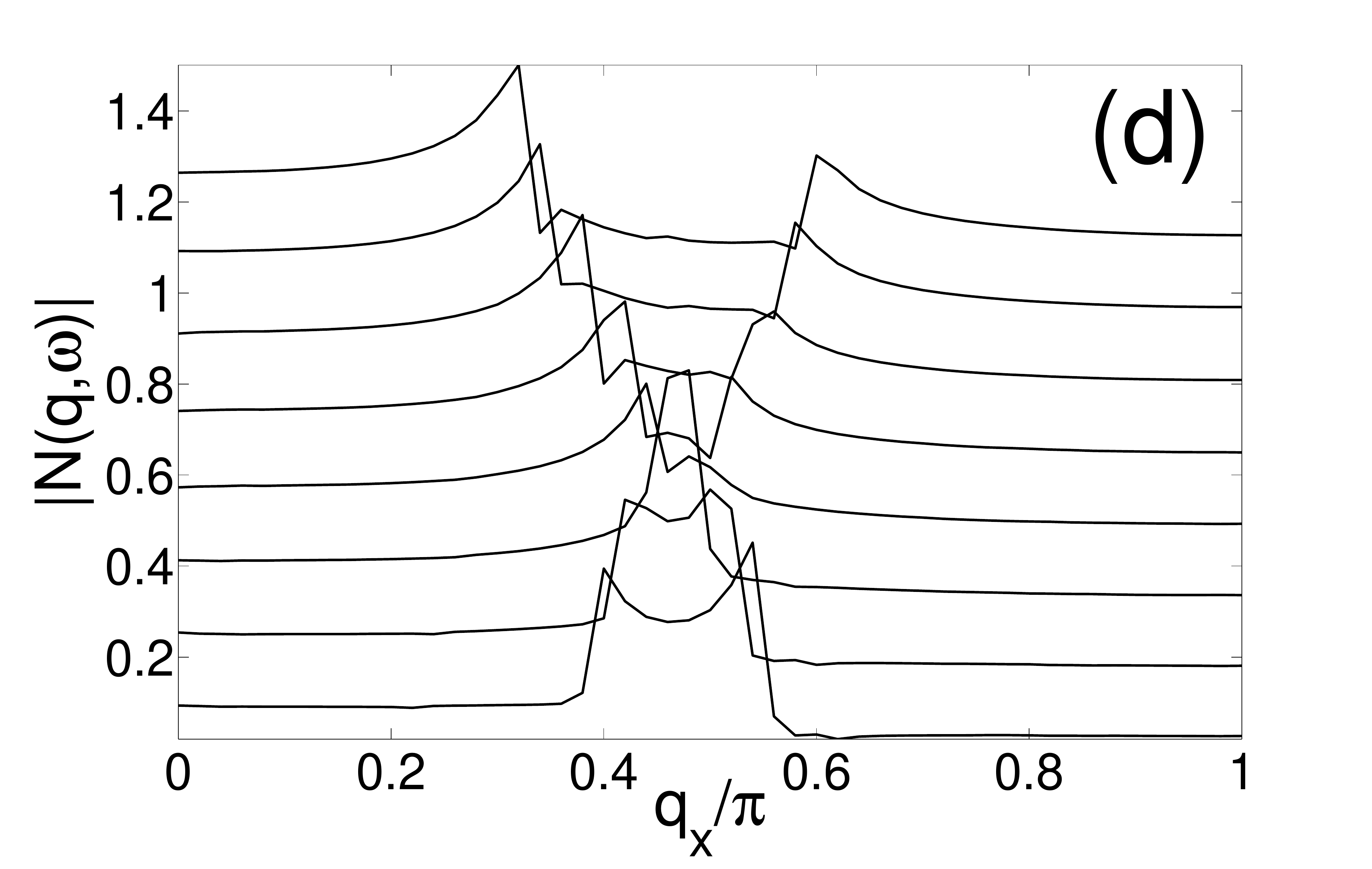}
\end{minipage}\caption{QPI line cuts along the (110) (a,b) and (100) (c,d)
directions for the pure AF case with $M=t$. Panels (a,c) [(b,d)]
correspond to positive [negative] energies at
$|\omega|/t=0.01,0.075,0.15,0.225,0.3,0.375,0.45,0.5$ (bottom to top).
For clarity the curves are displaced by 0.15. } \label{fig7}
\end{figure}

In Fig. \ref{fig7} we show line cuts along the nodal and antinodal
directions at both positive and negative energies. We can
understand the cusps in these line cuts from Fig. \ref{fig4}(d):
consider for example  the results presented in Fig. \ref{fig7}(a)
showing cuts at positive energies along the (110) direction. The
CCE clearly form ellipses centered around each nodal region as
seen from Fig. \ref{fig4}(d). Scattering can be inter-ellipse or
intra-ellipse, and leads to four cusps in agreement with Fig.
\ref{fig7}(a). Since at positive energies the CCE shrink with
increasing energy, the high momentum cusps resulting from
inter-ellipse scattering move up with energy, and eventually
merge. For the same reason, the low-momentum peaks resulting from
intra-ellipse scattering disperse downwards with increasing
energy. Since the negative energy CCE (not shown) expand with
energy, we find the opposite dispersion in this case as seen from
Fig. \ref{fig7}(b). From the cuts along the antinodal directions
shown in Figs. \ref{fig7}(c,d), we see that the QPI response is
actually strongest in this direction in agreement with the QPI
maps in Fig. \ref{fig5}

Finally we note that in samples with inhomogeneous coexisting regions of dSC and AF, one way to determine whether which contribution of the QPI signal originates mainly from the dSC or the AF regions, is to compare the result to the bias reversed QPI map. In the pure dSC state the same positions of $\q$-spots should appear but with different weight, whereas in the AF phase the arcs in the QPI maps have also dispersed resulting in new locations of the cusps in e.g. the line cuts shown in Fig. \ref{fig7}. Below, we study the simpler case of a homogeneous coexisting phase of dSC and AF long-range order.

\section{Coexisting superconducting and antiferromagnetic phase}

\begin{figure}[b]
\includegraphics[clip=true,width=1.0\columnwidth]{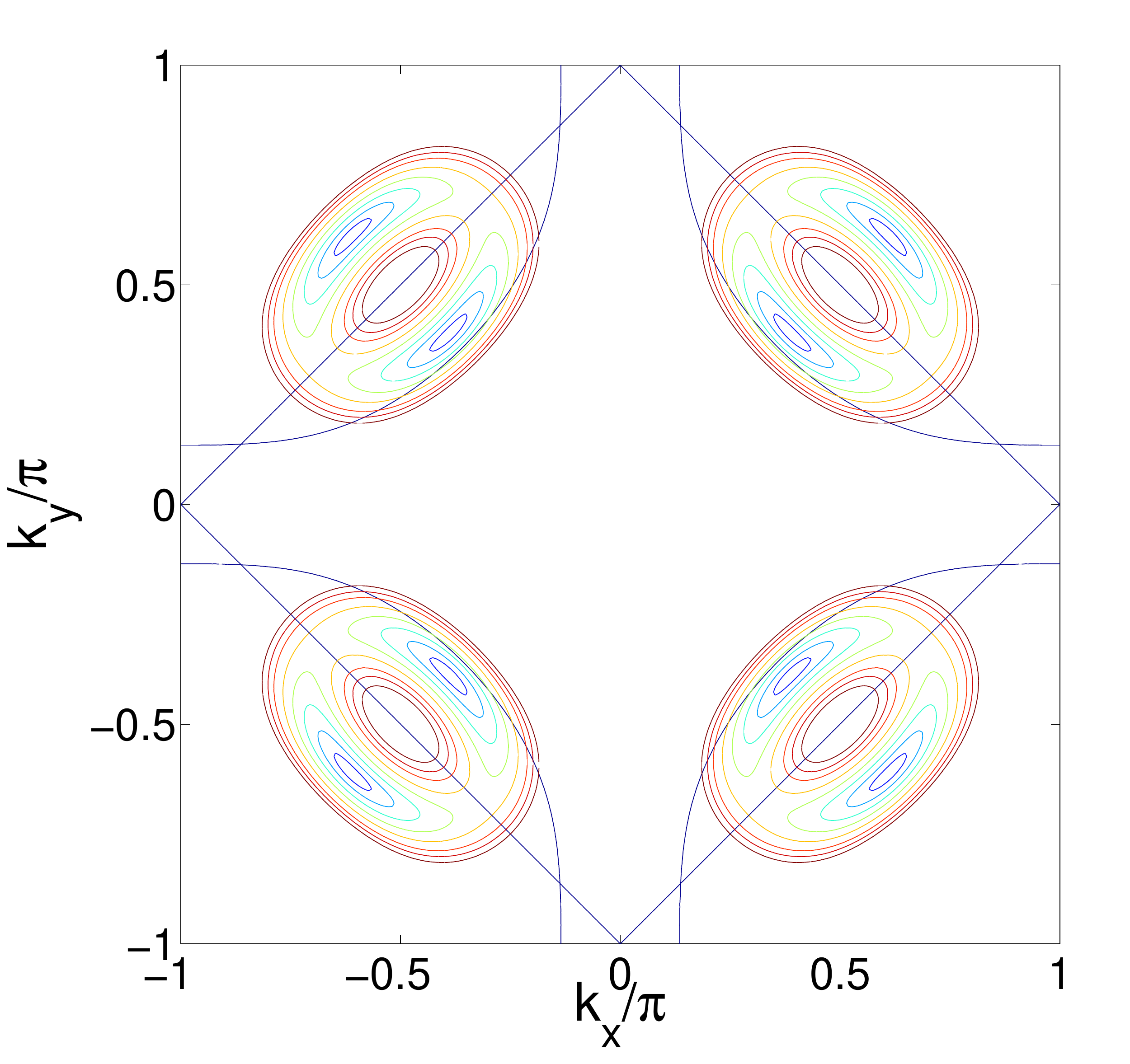}
\caption{(Color online) Contours of constant energy for the coexisting phase with $\Delta=0.6t$ and $M=t$ plotted at energies $\omega/t=0.075, 0.15, 0.225, 0.3, 0.375, 0.45, 0.5, 0.6$. The banana tips reach the AF zone boundary at $\omega=\Delta_0=0.33t$ when $M=t$. } \label{fig8}
\end{figure}

\begin{figure}[t]
\includegraphics[clip=true,height=1.5\columnwidth,width=1.0\columnwidth]{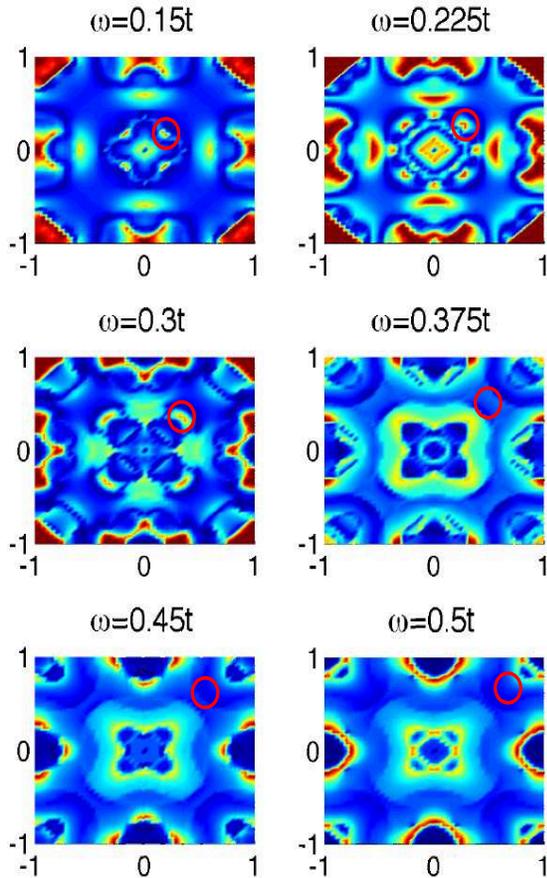}
\caption{(Color online) QPI maps versus $q_x/\pi$ and $q_y/\pi$ in the coexisting AF and dSC phase in the presence
of single point-like nonmagnetic impurity. Energies
$\omega$ are as shown in the Figure titles. This figure can be directly compared to Fig. \ref{fig2}.  Note the disappearance of the
spot-like feature corresponding to octet vector ${\bf q}_7$ above
the critical energy $\Delta_0=0.33t$.} \label{fig9}
\end{figure}

In the homogeneous coexisting state (AF+dSC) with $\Delta,M \neq 0$, the
$d$-wave gap collapses the arcs of spectral intensity at the Fermi
level shown in Fig. \ref{fig4}(a,b) to nodal points. This
robustness of the nodal points to antiferromagnetism away from
half-filling follows directly from the fact that ($\pi,\pi$) does
not nest the nodal points.\cite{eberg} Some representative CCE
for the coexisting phase are shown in Fig. \ref{fig8}. At low
energies the CCE are again reminiscent of dispersing bananas, but
now shifted off the normal state Fermi surface, and exist also in
the shadow band outside the RBZ.
This implies the existence of shadow banana tips and
shadow QPI peaks, whose weight will however generally be suppressed due
to coherence factors.

Once the banana tips reach the AF zone boundary, the CCE
become similar to the pure AF case. In Fig. \ref{fig9} we show the
results of the QPI patterns in the coexisting phase. This figure
can be directly compared to Fig. \ref{fig2}. We clearly see the
mixing of the AF and dSC QPI features and the importance of the
crossover scale set by the energy $\Delta_0$ where the CCE reach
the AF zone boundary: at $\omega<\Delta_0$ ($\omega>\Delta_0$) the
QPI is dominated by the pure dSC (AF) phase with the addition of
possible shadow band features resulting from scattering involving
the bananas outside the RBZ. Peaks generated by shadow bands are
particularly evident in the QPI image at $\omega=0.225t$ in Fig.
\ref{fig9}, and constitute a clear signature of the coexisting
phase. For BSCCO, however, it seems likely that the disorder is
simply too strong for these additional peaks to be observed [see
next section].  In Fig. \ref{fig9} we have circled the position of
the ${\bf q}_7$ peak from the pure dSC phase [see also Fig.
\ref{fig2}].  This peak appears to be extinguished when crossing
the AF zone boundary at $\omega=\Delta_0$. For the band parameter
and superconducting gap $\Delta=0.6t$ used in this paper, we have
$\Delta_0=0.45t$ for $M=0.5t$, and  $\Delta_0=0.33t$ for $M=t$.

\begin{figure}[]
\begin{minipage}{.49\columnwidth}
\includegraphics[clip=true,height=0.8\columnwidth,width=0.98\columnwidth]{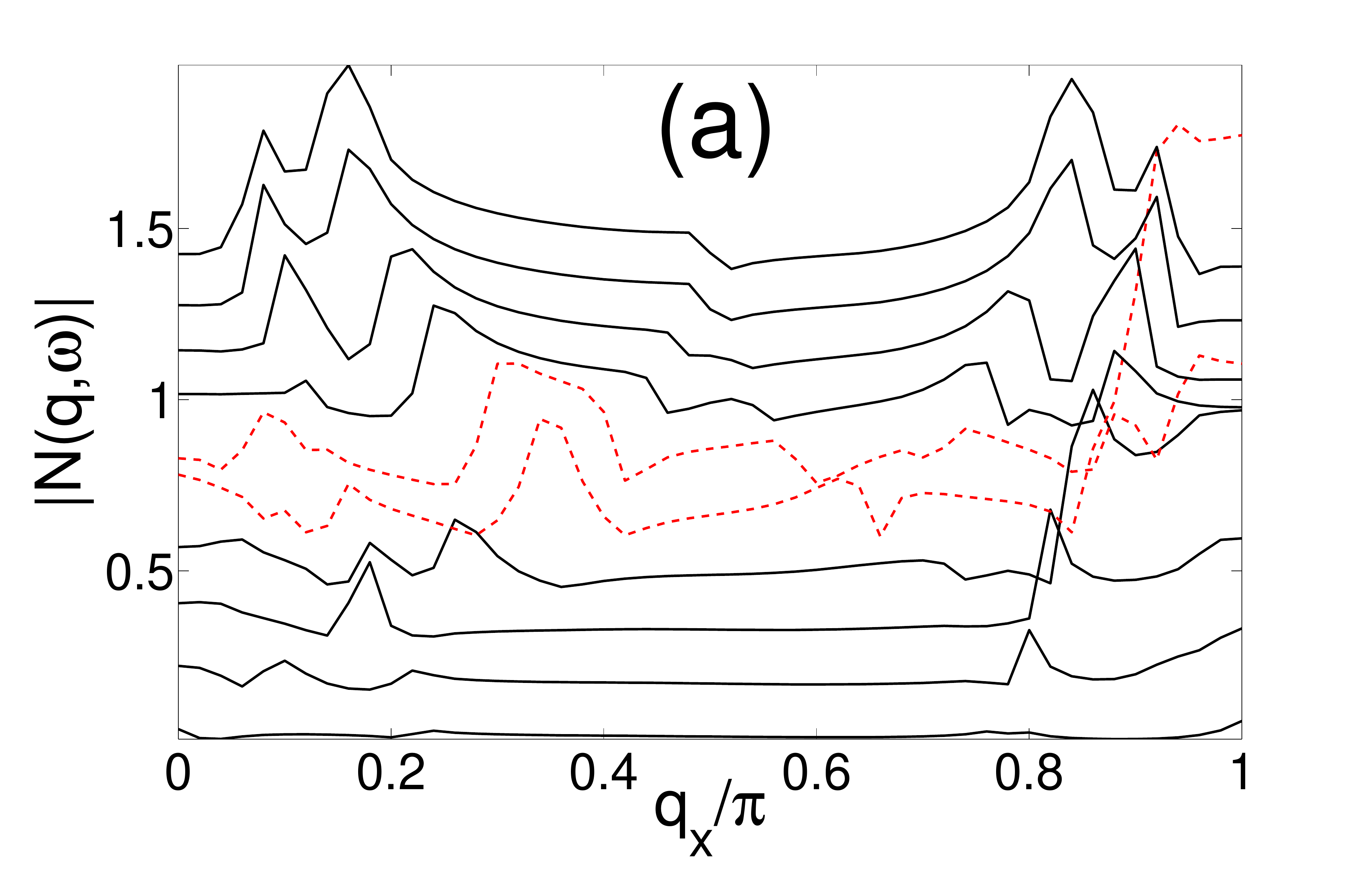}
\end{minipage}
\begin{minipage}{.49\columnwidth}
\includegraphics[clip=true,height=0.8\columnwidth,width=0.98\columnwidth]{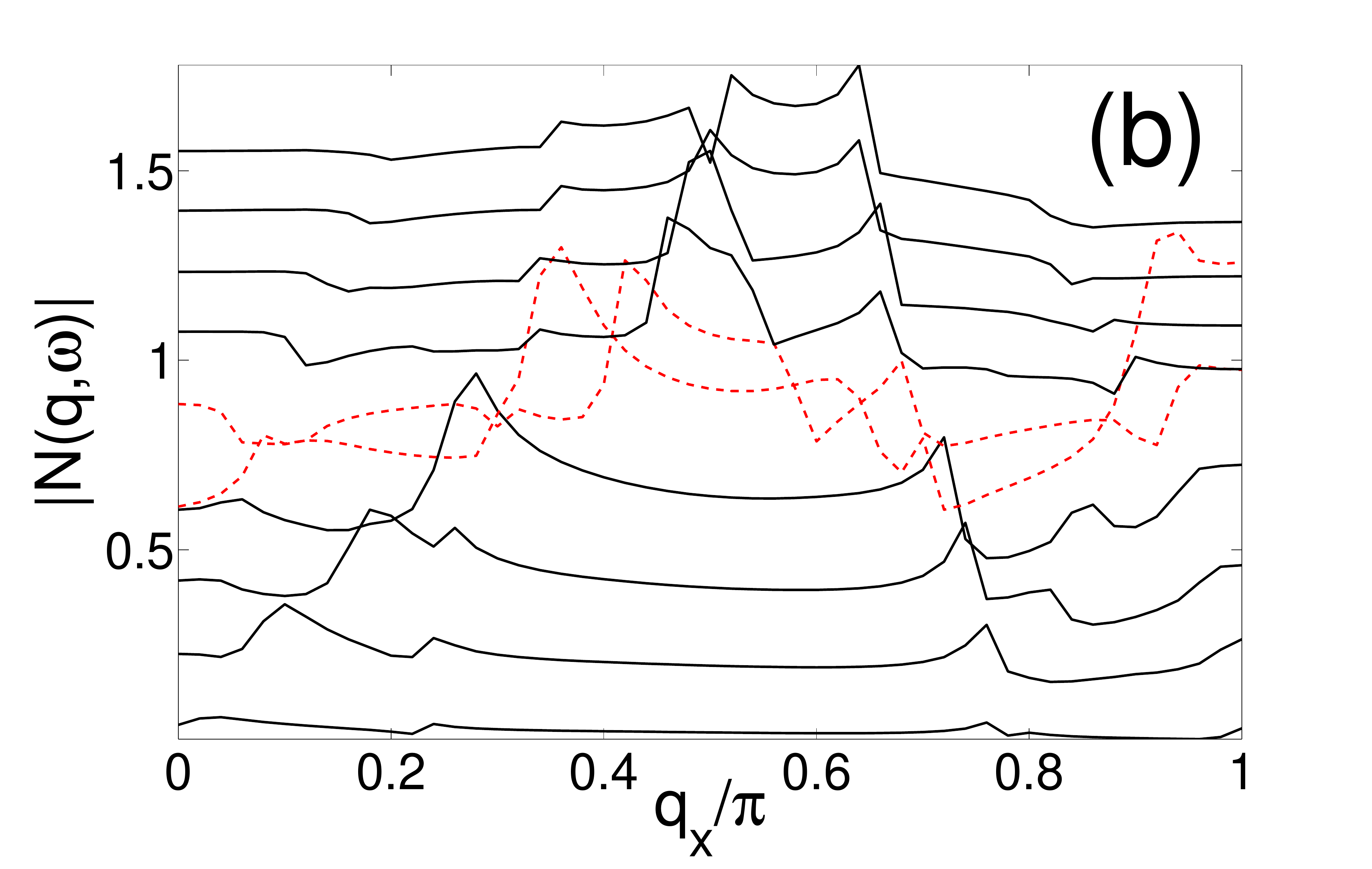}
\end{minipage}
\\
\begin{minipage}{.49\columnwidth}
\includegraphics[clip=true,height=0.8\columnwidth,width=0.98\columnwidth]{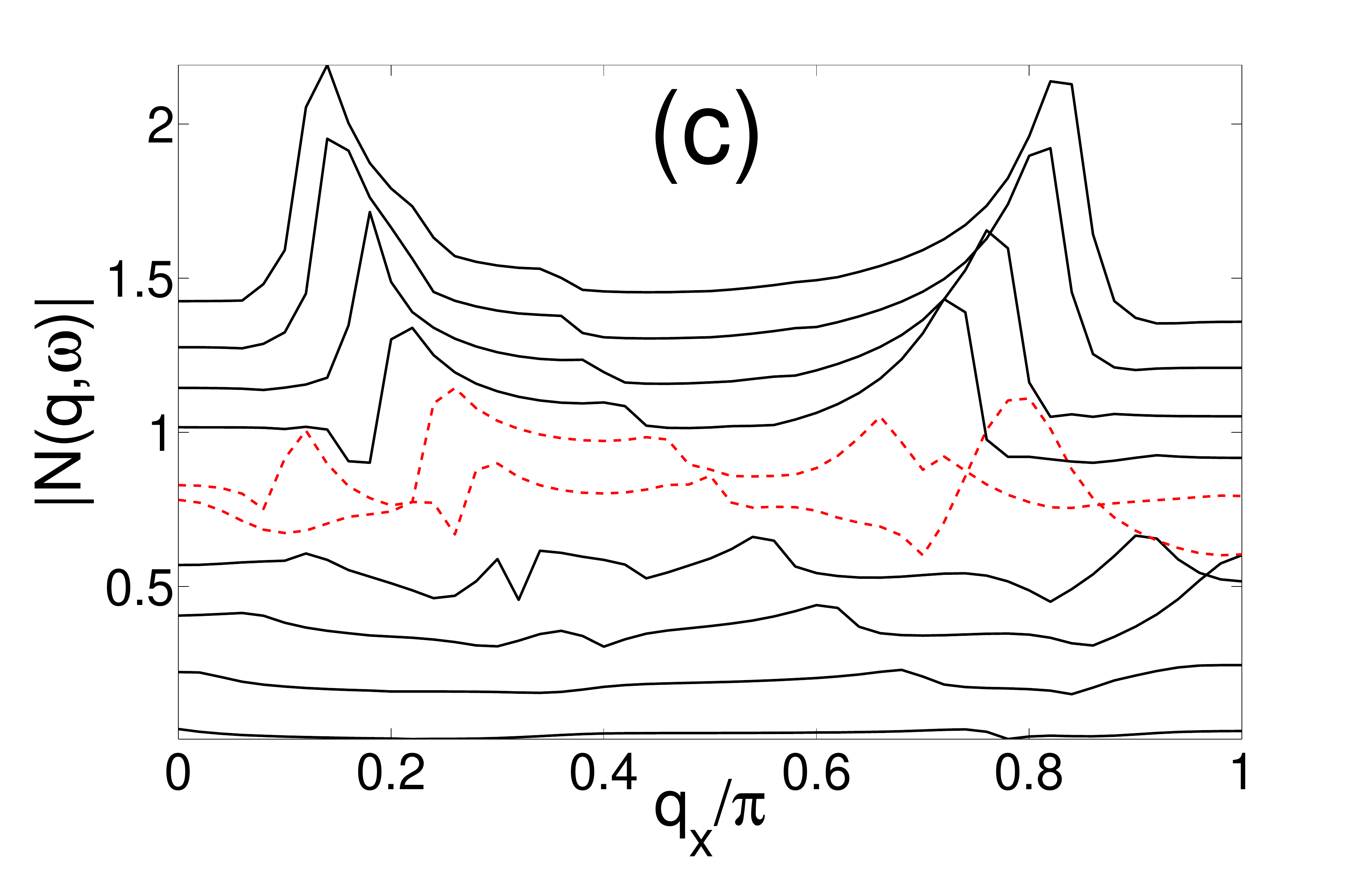}
\end{minipage}
\begin{minipage}{.49\columnwidth}
\includegraphics[clip=true,height=0.8\columnwidth,width=0.98\columnwidth]{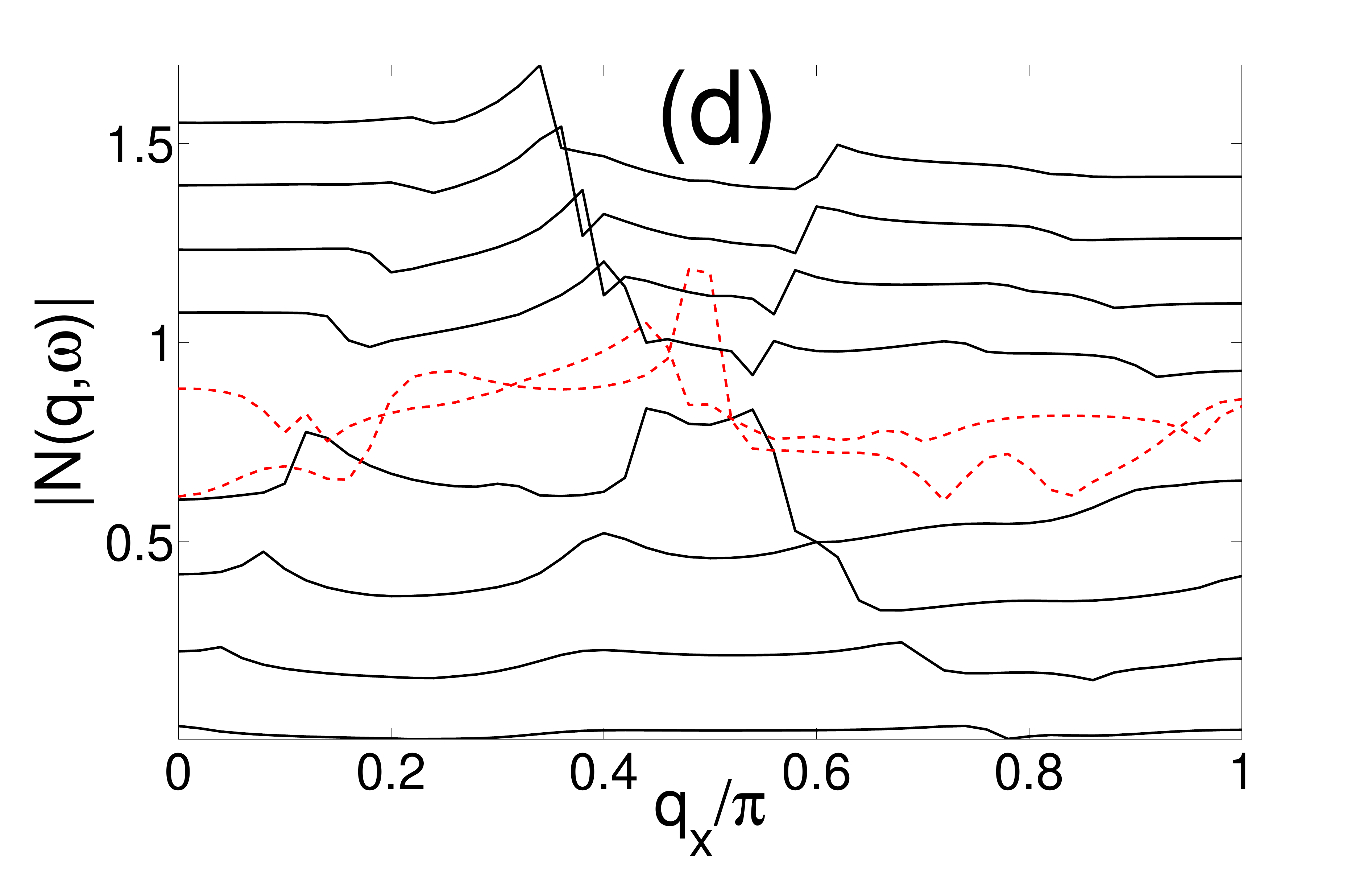}
\end{minipage}
\caption{(Color online) QPI line cuts along the (110) (a,b) and (100) (c,d) directions for the pure coexisting
AF and dSC phase with $M=t$ and $\Delta=0.6t$. Panels a,c (b,d) correspond to the following positive
 (negative) energies: $|\omega|/t=0.01,0.075,0.15,0.225,0.3,0.375,0.45,0.5,0.55,0.6$ (bottom to top).
 The line cuts close to the crossover region $\omega=\Delta_0$ are indicated by red dashed lines.
 For clarity the curves are displaced by 0.15.} \label{fig10}
\end{figure}

Figure \ref{fig10} shows the nodal and antinodal line cuts with
the red dashed lines indicating the crossover region near
$\Delta_0$. By comparison to Figs. \ref{fig3} and  \ref{fig7}, the
resemblance to the dSC and AF state is again clear for $\omega<\Delta_0$
and $\omega>\Delta_0$, respectively. Note that despite the
simplicity of the present model, it is evident from Figs.
\ref{fig9} and \ref{fig10} that at $\omega>\Delta_0$ the FT-STS
maps are dominated by slowly-dispersing  peaks in the antinodal
direction qualitatively similar to the experimental
data.\cite{kohsaka} These maxima disperse, however, more than
those in Ref. \onlinecite{kohsaka} and also appear to be more
arclike than the experimentally observed ${\bf q}_1^*$ and ${\bf
q_5^*}$ peaks.  This may be because our simple model assumes a
standard $d$-wave gap, whereas ARPES experiments reveal a gap
which varies little near the antinodal points.\cite{kanigel}  We
clearly also do not have in our model additional charge order leading to
the nondispersive QPI peaks.

\section{Effects of Disorder}

So far the discussion has focussed on pure phases, dSC, AF, and
AF+dSC in the presence of a single point-like nonmagnetic
impurity. Clearly this situation is rather simplistic and should be extended to
describe the QPI in a disordered short-range AF
cluster glass phase as relevant for LSCO, and possibly also BSCCO. One way to
model short-range AF correlations is by assuming that the AF
ordering vector ${\mathbf{Q}}$ follows a Lorentzian probability
distribution $p_Q$ peaked at $(\pi,\pi)$ and with a broadening
given by $\xi_{AF}^{-1}$, the inverse of the AF correlation
length. The main effects of disordering with such a distribution was studied in Ref. \onlinecite{singleton}: only the part
of the hole pockets which lie outside the RBZ is
shifted. This means, for example, that scattering events involving the outer shadow band bananas [see Fig. \ref{fig8}] will be smeared out by this kind of disorder averaging, rendering the low-energy FT-STS patterns virtually identical to those of the pure dSC state. Nevertheless, modeling the disorder by a $p_Q$ distribution appears questionable for the CSG phase since it does not properly treat the spatial inhomogeneity of the spin glass, and misses e.g. additional low-energy states existing at the boundary regions between magnetic and superconducting domains.\cite{andersen05} Therefore, it would be interesting to compare with quasiparticle interference patterns from more realistic
real-space disorder configurations similar to those produced in Refs. \onlinecite{alvarez05,atkinson05,robertson,delmaestro,andersen07,andersenkappa,alvarez08}. The results presented in this
paper should be helpful in providing an interpretation of results from future
more realistic disorder modeling.

\section{Conclusions}

We have studied  quasiparticle interference phenomena in a
dSC phase with coexisting short-range
AF order, as found, e.g.  in the cluster spin glass
phase of underdoped cuprates.  To do so, we have assumed that much
of the qualitative physics may be captured by studying the
analogous problem with long-range AF  order. In particular, we
have calculated the  QPI patterns arising from scattering from a
single point-like impurity in the case of a pure metallic AF phase and a
coexisting phase of both AF and dSC order. Due to different coherence factors in the AF and dSC phases, the QPI maps are dominated by peaks (arcs) in the dSC (AF) phase, respectively. In the case with coexisting order, low-energy quasiparticles propagate on contours which resemble the pure
superconductor, so dispersive, localized interference spots
similar to those predicted by the octet model for the pure
dSC system are recovered.  When the tips of these contours reach the AF zone face, however, the contours change abruptly to
those characteristic of the pure AF, and the dispersing low-energy localized interference peaks are then
extinguished, as reported by Kohsaka {\it et al.}\cite{kohsaka} We
expect that the remaining arc-like features at energies
$\omega>\Delta_0$ are suppressed by the short-range nature of the
true magnetic order in these systems.

The net result is a system where the low-energy quasiparticle
interference features resemble those of the optimally doped
superconducting samples, namely they disperse according to the
octet model.  For energies above the critical energy $\Delta_0$,
these localized spot-like features in momentum space
effectively disappear, as seen in experiment.  Although the
formation of AF long-range order itself may be viewed as a coherent multiple
scattering process, we see from the study of the coexisting AF-dSC
system that the quasiparticle states near the antinode are not
 destroyed by broadening in this process, but simply folded back.  Thus in the
realistic situation with short-range AF order, some scattering
from AF modulations will occur, but still well-defined
quasiparticle peaks should remain, as seen in ARPES.  This
resolves an apparent paradox in the comparison of the two
experimental techniques in their view of the antinodal
quasiparticles.   It is the localized QPI ${\bf q}$-space
structures due to these states which are destroyed, not the states
themselves.

We emphasize that the formalism presented in this paper is
essentially identical for any competing order scenario with ordering wavevector near $(\pi,\pi)$.  We believe that the bulk of the
experimental evidence support the identification of the competing
phase with short-range incommensurate AF order or quasistatic fluctuations from
incipient order, as observed in $\mu$SR.  We have therefore
discussed the extinction of QPI above a critical energy in this
context, but other explanations with a similar structure may be
possible.

In addition, we note that realistic simulations 
of the QPI patterns require a more sophisticated description
of disorder than that adopted here.  It is thought that 
the potential produced by a single defect has components in
at least the screened Coulomb and pairing channels, and some
account of the latter is known to be necessary to reproduce
the correlations between dopant position and gap size observed in
the BSCCO system.\cite{NAMH05}  We have neglected these
details here, but they are discussed e.g. in Ref. \onlinecite{nunner}. 
Nevertheless, the patterns produced by our single potential scatterer
were found to be sufficient to describe the basic phenomenology
 of Kohsaka {\it et al.}\cite{kohsaka} for the octet $\q_7$ peak. It will be 
interesting to see if more sophisticated
simulations can also reproduce the behavior of the other octet
vectors and their weights correctly.

One aspect of the current picture which remains unclear is the
extent to which static order is required.  The QPI
extinction phenomenon observed by Kohsaka {\it et al.}\cite{kohsaka} is observed in BSCCO samples with doping levels 6-19\%, i.e. including optimal doping where recent neutron experiments have claimed a 32 meV spin
gap.\cite{fauque} Still, near ($\pi,\pi$), some
intensity remains below this energy, and  $\mu$SR continues to
indicate frozen magnetic order even in these samples at low
temperatures.\cite{panagopoulos} It may be that the magnetism in this system is simply
too disordered to be seen by current neutron experiments. A more
sophisticated treatment of short-range order is required to
address these questions.

\section{Acknowledgement}

The authors acknowledge valuable discussions with J. C. Davis, Z.-X.
Shen, and M. Vojta. B.~M.~A. acknowledges support from the Villum
Kann Rasmussen foundation, and P.~J.~H. from DOE-BES DE-FG02-05ER46236.


\begin{thebibliography}{00}

\bibitem{Damascelli03}  A. Damascelli, Z. Hussain, and Z.-X. Shen, Rev. Mod. Phys. {\bf 75}, 473 (2003).
%
\bibitem{FuLee06} H. Fu. and D.-H. Lee, Phys. Rev. B {\bf 74}, 174513 (2006).
%
\bibitem{Hoffman1} J. E. Hoffman, K. McElroy, D.-H. Lee, K. M. Lang, H. Eisaki, S. Uchida, and J. C. Davis, Science {\bf 297}, 1148 (2002).
%
\bibitem{Howald} C. Howald, H. Eisaki, N. Kaneko, M. Greven, and A. Kapitulnik, Phys. Rev. B {\bf 67}, 014533 (2003).
%
\bibitem{Hoffman2} K. McElroy, R. W. Simmonds, J. E. Hoffman, D.-H. Lee, J. Orenstein, H. Eisaki, S. Uchida, J. C. Davis, Nature (London) {\bf 422}, 592 (2003).
%
\bibitem{vershinen} M. Vershinin, S. Misra, S. Ono, Y. Abe, Y. Ando, and A. Yazdani, Science {\bf 303}, 1995 (2004).
%
\bibitem{mcelroy1} K. McElroy, D.-H. Lee, J. E. Hoffman, K. M. Lang, J. Lee, E. W. Hudson, H. Eisaki, S. Uchida, and J. C. Davis, Phys. Rev. Lett. {\bf 94}, 197005 (2005).
%
\bibitem{hashimoto} A. Hashimoto, N. Momono, M. Oda, and M. Ido, Phys. Rev. B {\bf 74}, 064508 (2006).
%
\bibitem{WangLee} Q.-H. Wang and D.-H. Lee, Phys. Rev. B {\bf 67}, 020511 (2003).
%
\bibitem{Sprunger} P. T. Sprunger, L. Petersen, E. W. Plummer, E. L\ae gsgaard, and F. Besenbacher, Science {\bf 275}, 1764 (1997).
%
\bibitem{Tingfourier} D. Zhang and C. S. Ting, Phys. Rev. B {\bf 67}, 100506 (2003).
%
\bibitem{Franzold} T. Pereg-Barnea and M. Franz, Phys. Rev. B {\bf 68}, 180506 (2003).
%
\bibitem{capriotti} L. Capriotti, D. J. Scalapino, and R. D. Sedgewick, Phys. Rev. B {\bf 68}, 014508 (2003).
%
\bibitem{ZAH04} L.-Y. Zhu, W. A. Atkinson, and P. J. Hirschfeld, Phys. Rev. B {\bf 69}, 060503 (2004).
%
\bibitem{misra} S. Misra, M. Vershinen, P. Phillips, and A. Yazdani, Phys. Rev. B {\bf 70}, 220503 (2004).
%
\bibitem{mcelroyscience} K. McElroy, J. Lee, J. A. Slezak, D.-H. Lee, H. Eisaki, S. Uchida, and J. C. Davis, Science {\bf 309}, 1048 (2005).
%
\bibitem{NAMH05} T. S. Nunner, B. M. Andersen, A. Melikyan, and P. J. Hirschfeld, Phys. Rev. Lett. {\bf 95}, 177003 (2005).
%
\bibitem{andersen06} B. M. Andersen, A. Melikyan, T.  S. Nunner, and P. J. Hirschfeld, Phys. Rev. B {\bf 74}, 060501 (2006).
%
\bibitem{nunner} T. S. Nunner, W. Chen, B.  M. Andersen, A. Melikyan, and P. J. Hirschfeld, Phys. Rev. B {\bf 73}, 104511 (2006).
%
\bibitem{DellAnna} L. Dell'Anna, J. Lorenzana, M. Capone, C. Castellani, and M. Grilli, Phys. Rev. B {\bf 71}, 064518 (2005).
%
\bibitem{ChengSu} M. Cheng and W. P. Su, Phys. Rev. B {\bf 72}, 094512 (2005).
%
\bibitem{liu} Y. H. Liu, K. Takeyama, T. Kurosawa, N. Momono, M. Oda, and M. Ido, Phys. Rev. B {\bf 75}, 212507 (2007).
%
\bibitem{hanaguri1} T. Hanaguri, C. Lupien, Y. Kohsaka, D.-H. Lee, M. Azuma, M. Takano, H. Takagi, and J. C. Davis, Nature (London) {\bf 430}, 1001 (2004).
%
\bibitem{wise} W. D. Wise, M. C. Boyer, K. Chatterjee, T. Kondo, T. Takeuchi, H. Ikuta, Y. Wang, and E. W. Hudson, Nature Phys. {\bf 4}, 696 (2008).
%
\bibitem{sachdev} A. Polkovnikov, M. Vojta, and S. Sachdev, Phys. Rev. B {\bf 65}, 220509 (2002); Physics C {\bf 388-389}, 19 (2003).
%
\bibitem{chenyeh} C.-T. Chen and N.-C. Yeh, Phys. Rev. B {\bf 68}, 220505 (2003).
%
\bibitem{podolsky} D. Podolsky, E. Demler, K. Damle, and B. I. Halperin, Phys. Rev. B {\bf 67}, 094514 (2003).
%
\bibitem{AHB03} B. M. Andersen, H. Bruus, and P. Hedeg\aa rd, Phys. Rev. B {\bf 67}, 134528 (2003).
%
\bibitem{linda} L. Udby, B. M. Andersen, and P. Hedeg\aa rd, Phys. Rev. B {\bf 73}, 224510 (2006).
%
\bibitem{DHLee} J.-X. Li, C.-Q. Wu, and D.-H. Lee, Phys. Rev. B {\bf 74}, 184515 (2006).
%
\bibitem{brown} S. E. Brown, E. Fradkin, and S. A. Kivelson, Phys. Rev. B {\bf 71}, 224512 (2005).
%
\bibitem{ghosal}  A. Ghosal, A. Kopp, and S. Chakravarty, Phys. Rev. B {\bf 72}, 220502 (2005).
%
\bibitem{chatterjee} U. Chatterjee, M. Shi, A. Kaminski, A. Kanigel, H. M. Fretwell, K. Terashima, T. Takahashi, S. Rosenkranz, Z. Z. Li, H. Raffy, A. Santander-Syro, K. Kadowaki, M. R. Norman, M. Randeria, and J. C. Campuzano, Phys. Rev. Lett. {\bf 96}, 107006 (2006).
%
\bibitem{bascones} E. Bascones and B. Valenzuela, Phys. Rev. B {\bf 77}, 024527 (2008).
%
\bibitem{hanaguri2} T. Hanaguri, Y. Kohsaka, J. C. Davis, C. Lupien, I. Yamada, M.
Azuma, M. Takano, K. Ohishi, M. Ono, and H. Takagi, Nature Phys.  {\bf 3}, 865 (2007).
%
\bibitem{kohsaka}Y. Kohsaka, C. Taylor, P. Wahl, A. Schmidt, J. Lee, K. Fujita, J. Alldredge, J. Lee, K. McElroy, H. Eisaki, S. Uchida, D.-H. Lee, and J. C. Davis, Nature (London) {\bf 454}, 1072 (2008).
%
\bibitem{hanaguri3}T. Hanaguri, Y. Kohsaka, M. Ono, M. Maltseva, P. Coleman, I. Yamada, M. Azuma, M. Takano, K. Ohishi, and H. Takagi (unpublished).
%
\bibitem{graserprb07} S. Graser, P. J. Hirschfeld, and D. J. Scalapino,  Phys. Rev. B {\bf 77}, 184504 (2008).
%
\bibitem{eschrigreview} M. Eschrig, Adv. Phys. {\bf 55}, 47 (2006).
%
\bibitem{Dahmetal1} T. Dahm and L. Tewordt, Phys. Rev. Lett. {\bf 74}, 793 (1995).
%
\bibitem{Quinlan96} S. Quinlan, P. J. Hirschfeld, and D. J. Scalapino, Phys. Rev. B {\bf 53}, 8575 (1996).
%
\bibitem{Dahmetal2}  T. Dahm, P. J. Hirschfeld, D. J. Scalapino, and L.-Y. Zhu,  Phys. Rev. B {\bf 72}, 214512 (2005).
%
\bibitem{campuzano} J. C. Campuzano, H. Ding, M. R. Norman, H. M. Fretwell, M. Randeria, A. Kaminski, J. Mesot, T. Takeuchi, T. Sato, T. Yokoya, T. Takahashi, T. Mochiku, K. Kadowaki, P. Guptasarma, D. G. Hinks, Z. Konstantinovic, Z. Z. Li, and H. Raffy, Phys. Rev. Lett. {\bf 83}, 3709 (1999).
%
\bibitem{shi} M. Shi, A. Bendounan, E. Razzoli, S. Rosenkranz, M. R. Norman, J. C. Campuzano, J. Chang, M. Mansson, Y. Sassa, T. Claesson, O. Tjernberg, L. Patthey, N. Momono, M. Oda, M. Ido, S. Guerrero, C. Mudry, and J. Mesot, arXiv:0810.0292v1.
%
\bibitem{Tranquadareview} J. Tranquada in: {\it Handbook of
High Temperature Superconductivity: Theory and Experiment}, editor J. R. Schrieffer, associate editor J. S. Brooks (Springer Verlag, Hamburg, 2007).
%
\bibitem{fujita} M. Fujita, H. Goka, K. Yamada, J. M. Tranquada, and L. P. Regnault, Phys. Rev. B {\bf 70}, 104517 (2004).
%
\bibitem{julien} M.-H. Julien, Physica B {\bf 329-333}, 693  (2003).
%
\bibitem{panagopoulos} C. Panagopoulos  J. L. Tallon, B. D. Rainford, J. R. Cooper, C. A Scott, and T. Xiang, Solid State Comm. {\bf 126}, 47 (2003).
%
\bibitem{xjzhou} X. J. Zhou, T. Yoshida, D.-H. Lee, W. L. Yang, V. Brouet, F. Zhou, W. X. Ti, J. W. Xiong, Z. X. Zhao, T. Sasagawa, T. Kakeshita, H. Eisaki, S. Uchida, A. Fujimori, Z. Hussain, and Z.-X. Shen, Phys. Rev. Lett. {\bf 92}, 187001 (2004).
%
\bibitem{kivelson}  S. A. Kivelson, I. P. Bindloss, E. Fradkin, V. Oganesyan, J. M. Tranquada, A. Kapitulnik, and C. Howald, Rev. Mod. Phys. {\bf 75}, 1201 (2003).
 %
\bibitem{alvarez05} G. Alvarez,  M. Mayr, A. Moreo, and E. Dagotto, Phys. Rev. B {\bf 71}, 014514 (2005); M. Mayr, G. Alvarez, A. Moreo, and E. Dagotto, Phys. Rev. B {\bf 73}, 014509 (2006).
%
\bibitem{atkinson05} W. A. Atkinson, Phys. Rev. B {\bf 71}, 024516 (2005).
%
\bibitem{robertson} J. A. Robertson, S. A. Kivelson, E. Fradkin, A. C. Fang, and A. Kapitulnik, Phys. Rev. B {\bf 74}, 134507 (2006).
%
\bibitem{delmaestro} A. Del Maestro, B. Rosenow, and S. Sachdev, Phys. Rev. B {\bf 74}, 024520 (2006).
%
\bibitem{kaul} M. Vojta, T. Vojta, and R. K. Kaul, Phys. Rev. Lett. {\bf 97}, 097001 (2006).
%
\bibitem{andersen07} B. M. Andersen, P. J. Hirschfeld, A. P. Kampf, and M. Schmid, Phys. Rev. Lett. {\bf 99}, 147002 (2007).
%
\bibitem{ZWang} C. Li, S. Zhou, and Z. Wang, Phys. Rev. B {\bf 73}, 060501 (2006).
%
\bibitem{seo} K. Seo, H.-D. Chen, and J. Hu, Phys. Rev. B {\bf 76}, 020511 (2007); Phys. Rev. B {\bf 78}, 094510 (2008).
%
\bibitem{vojta08} M. Vojta and O. R\"osch, Phys. Rev. B {\bf 77}, 094504 (2008).
%
\bibitem{atkinson07} W. A. Atkinson, Phys. Rev. B {\bf 75}, 024510 (2007).
%
\bibitem{andersenkappa} B. M. Andersen and P. J. Hirschfeld, Phys. Rev. Lett. {\bf 100}, 257003 (2008).
%
%
\bibitem{alvarez08} G. Alvarez and E. Dagotto, Phys. Rev. Lett. {\bf 101}, 177001 (2008).
%
\bibitem{andersenDresden} B. M. Andersen and P. J. Hirschfeld, Physica (Amsterdam) {\bf 460C}, 744
(2007).
%
\bibitem{lake} B. Lake, H. M. R\o nnow, N. B. Christensen, G. Aeppli, K. Lefmann, D. F. McMorrow, P. Vorderwisch, P. Smeibidl, N. Mangkorntong, T. Sasagawa, M. Nohara, H. Takagi, and T. E. Mason, Nature (London) {\bf 415}, 299 (2002).
%
\bibitem{bena} C. Bena, S. Chakravarty, J. Hu, and C. Nayak, Phys. Rev. B {\bf 69}, 134517 (2004).
%
\bibitem{chakravarty} S. Chakravarty, C. Nayak, and S. Tewari,  Phys. Rev. B {\bf 68}, 100504
(2003).
%
\bibitem{singleton}N. Harrison, R. D. McDonald, and J. Singleton, Phys. Rev. Lett. {\bf 99}, 206406 (2007).
%
\bibitem{eberg} E. Berg, C.-C. Chen, and S. A. Kivelson, Phys. Rev. Lett. {\bf 100}, 027003 (2008).
%
\bibitem{kanigel} A. Kanigel, M. R. Norman, M. Randeria, U. Chatterjee, S. Souma, A. Kaminski, H. M. Fretwell, S. Rosenkranz, M. Shi, T. Sato, T. Takahashi, Z. Z. Li, H. Raffy, K. Kadowaki, D. Hinks, L. Ozyuzer, and J. C. Campuzano, Nature Phys.  {\bf 2}, 447 (2006).
%
\bibitem{andersen05} B. M. Andersen, I. V. Bobkova, P. J. Hirschfeld, and Yu. S. Barash, Phys. Rev. B {\bf 72}, 184510 (2005).
%
\bibitem{fauque} B. Fauqu\'e, Y. Sidis, L. Capogna, A. Ivanov, K. Hradil, C. Ulrich, A.I. Rykov, B. Keimer, and P. Bourges, Phys. Rev. B {\bf 76}, 214512 (2007).
%
\end{thebibliography}
\end{document}